\pdfoutput=1
\documentclass[fleqn,usenatbib]{mnras}
\usepackage{txfonts}

\usepackage[T1]{fontenc}
\usepackage{graphicx}

\usepackage{graphicx}	
\usepackage{longtable}

\volume{{\rm in press}}

\title[QPPs in Stellar Flares]{Statistical Properties of Quasi-Periodic Pulsations in White-Light Flares Observed With \emph{Kepler}}

\author[C. E. Pugh et al.]{
C. E. Pugh,$^{1}$\thanks{E-mail: c.e.pugh@warwick.ac.uk}
D. J. Armstrong,$^{1,2}$
V. M. Nakariakov,$^{1,3}$
A.-M. Broomhall,$^{1,4}$
\\
$^{1}$Department of Physics, University of Warwick, Coventry, CV4 7AL, UK\\
$^{2}$ARC, School of Mathematics \& Physics, Queen's University Belfast, Belfast, BT7 1NN, UK\\
$^{3}$Central Astronomical Observatory at Pulkovo of RAS, St Petersburg, 196140, Russia\\
$^{4}$Institute of Advanced Study, University of Warwick, Coventry, CV4 7HS, UK
}

\date{Accepted XXX. Received YYY; in original form ZZZ}

\pubyear{2016}

\begin{document}
\label{firstpage}
\pagerange{\pageref{firstpage}--\pageref{lastpage}}
\maketitle

\begin{abstract}

We embark on a study of quasi-periodic pulsations (QPPs) in the decay phase of white-light stellar flares observed by \emph{Kepler}. Out of the 1439 flares on 216 different stars detected in the short-cadence data using an automated search, 56 flares are found to have pronounced QPP-like signatures in the light curve, of which 11 have stable decaying oscillations. No correlation is found between the QPP period and the stellar temperature, radius, rotation period and surface gravity, suggesting that the QPPs are independent of global stellar parameters. Hence they are likely to be the result of processes occurring in the local environment. There is also no significant correlation between the QPP period and flare energy, however there is evidence that the period scales with the QPP decay time for the Gaussian damping scenario, but not to a significant degree for the exponentially damped case. This same scaling has been observed for MHD oscillations on the Sun, suggesting that they could be the cause of the QPPs in those flares. Scaling laws of the flare energy are also investigated, supporting previous reports of a strong correlation between the flare energy and stellar temperature/radius. A negative correlation between the flare energy and stellar surface gravity is also found.

\end{abstract}

\begin{keywords}
stars: activity -- stars: coronae -- stars: flare -- stars: oscillations -- Sun: flares -- Sun: oscillations
\end{keywords}



\section{INTRODUCTION}

Pulsations in solar flares were first detected by \citet{1969ApJ...155L.117P}, and with the development of increasingly high-precision instruments that observe the Sun, it emerged that these quasi-periodic pulsations (QPPs) of the energy release are a common feature of flares \citep[e.g.][]{2010SoPh..267..329K, 2015SoPh..290.3625S}. Properties of the QPPs relate to the plasma parameters in the flaring region, and to the physical processes in operation. There are two categories of mechanisms that have been proposed to explain the origin of QPPs, which are the ``magnetic dripping" mechanisms and magnetohydrodynamic (MHD) oscillations \citep{2009SSRv..149..119N}. The magnetic dripping mechanisms are based on the idea that a continuous supply of free magnetic energy could cause magnetic reconnection to repetitively occur each time a threshold energy is surpassed. This quasi-periodic self-oscillatory regime of spontaneous magnetic reconnection would result in quasi-periodic modulation of particle acceleration and the rate of energy release. The observed periodicity would hence relate to the steady inflow of magnetic energy. Examples of this regime have been found in numerical simulations by \citet{2000A&A...360..715K, 2009A&A...494..329M, 2012A&A...548A..98M}. QPPs caused by MHD oscillations could involve either oscillations of the flaring region itself, or MHD oscillations of a nearby structure. In the first case, one possibility is that the variation of parameters of the flaring plasma (such as the magnetic field and plasma density) directly modulates the radiation emission due to the gyrosynchrotron mechanism or bremsstrahlung, and the other is periodic modulation of the particle acceleration resulting in modulation of the emission \citep[e.g.][]{2008PhyU...51.1123Z}. The latter case may be considered as a periodically triggered regime of magnetic reconnection, where the fast or slow magnetoacoustic oscillations leak from the oscillating structure and approach the flaring site, resulting in, for example, plasma micro-instabilities and hence anomalous resistivity, which could periodically trigger magnetic reconnection \citep{2006A&A...452..343N, 2006SoPh..238..313C}.

The first observation of pulsations in a stellar flare was made by \citet{1974A&A....32..337R}, and since then occasional observations of QPPs in different stars have been made in the optical \citep{2003A&A...403.1101M}, ultraviolet \citep{2006A&A...458..921W}, microwave \citep{2004AstL...30..319Z} and X-ray \citep{2005A&A...436.1041M, 2009ApJ...697L.153P} wavebands. \citet{2015ApJ...813L...5P} found a rare example of a stellar superflare with multiple statistically significant periodicities, on a star observed by NASA's \emph{Kepler} mission. The properties of the QPPs were consistent with multiple periodicities observed in some solar flares, and with MHD oscillations being the cause, suggesting that the same physical processes operate in both solar flares and superflares on cool main sequence stars.

The \emph{Kepler} mission made white-light observations of around 150,000 astrophysical targets between 2009 and 2013. Despite its primary purpose being to allow the detection of (preferably habitable) exoplanets, it is also proving to be a valuable resource for the study of stellar flares. In addition, the availability of data with a cadence of one minute makes \emph{Kepler} suitable for studying QPPs with periods greater than a few minutes, allowing for the sample of known stellar flares with QPPs to be increased substantially. The first 7 flares in the \emph{Kepler} data with evidence of QPPs were reported by \citet{2015MNRAS.450..956B}, with period ranging from 4.8 to 14 minutes. \citet{2015EP&S...67...59M} also noted that several flares showed QPP-like signatures in their study of superflares on solar-type stars, and \citet{2014ApJ...797..122D} classified 15.5\% of flares detected in the \emph{Kepler} light curves of the highly active star GJ 1243 (KIC 9726699) as being complex, due to them having multiple peaks. In addition to this, \citet{2014ApJ...797..122D} found that a broken power-law fitted the distribution of flare durations very well, but this broken power-law model could not fully reproduce the observed fraction of complex flares as a function of duration. They concluded that this could be due to some of the apparent complex flares being the superposition of multiple independent flares. Also, they fitted complex flares with a model based on the superposition of multiple flare shapes. Complex flares which are well fitted by this model and whose multiple peaks are not periodic are likely to be sympathetic flares.

Several studies have shown that hotter stars flare less frequently \citep[e.g.][]{2014ApJ...792...67C}. Other results have also suggested that hotter, more luminous stars produce flares with higher energies; for example, \citet{1984ApJS...54..375P} found a correlation between the cumulative flare energy distribution and stellar temperature using data from 7 stars. A large-scale study of M dwarfs also found that the higher luminosity flares were less likely to occur on the cooler, redder stars \citep{2009AJ....138..633K}: a result which was supported by \citet{2011AJ....141...50W} using \emph{Kepler} Quarter 1 observations of cool stars. The complete data set from \emph{Kepler} allows us to investigate this possible dependency of the flare energy on stellar parameters further, and recently \citet{2015MNRAS.447.2714B} found a strong positive correlation between flare energy and stellar radius/luminosity. Since it is thought that the magnetic field of early-type stars tends to be weaker than that of late-type stars, this result suggested that the flaring active region size scales with the size of the star, as $E \approx L^3B^2/8\pi$, where $E$ is the total magnetic energy stored in the active region, $L$ is the characteristic size of the active region, and $B$ is the magnetic field strength. \citet{2015MNRAS.447.2714B} also found that the flare energy scales with the host star radius cubed, much like how the energy stored in an active region scales with $L^3$, providing further support to this idea. Hence it was suggested that the relationship between flare energy and stellar radius could be used to constrain the magnetic field strength of the active region that produced the flare. This approach should be treated with caution, however, as it is the free magnetic energy stored in an active region which gives an upper limit on the possible flare energy, rather than the total magnetic energy.

In this paper we present the largest collection of stellar flares exhibiting QPPs studied to date, all observed by the \emph{Kepler} mission. Section~\ref{sec:method} gives details about the data and analysis methods used. We derive parameters of the QPPs and investigate whether there is a relationship between the QPP period and stellar temperature, radius, rotation period and surface gravity in Section~\ref{sec:results}. The distribution of QPP periods is also studied, and any dependencies on other properties of the flare, namely the total energy and QPP decay time, are checked for. A summary is given in Section~\ref{sec:conc}, and in Appendix~\ref{sec:A} possible correlations between the flare energy and stellar parameters are explored.

\section{DATA AND ANALYSIS}
\label{sec:method}

\subsection{Identifying decaying oscillations in the flares}
\label{sec:id}

Stars classified as A- to M-type observed in short-cadence mode by \emph{Kepler} were searched for flares using an automated algorithm. This method involved removing long-duration trends in the simple aperture photometry (SAP) light curves by smoothing them over an interval of 100 minutes, and subtracting the smoothed version from the original light curve. Any time locations where more than two consecutive data points had values greater than 4.5 times the standard deviation of the de-trended light curve were then flagged. Light curves containing candidate flares were checked by eye for the characteristic flare shape --- a rapid rise in flux followed by a more gradual decline --- and any flares showing potential signs of QPPs were analysed using the wavelet and autocorrelation techniques. We chose to focus on QPPs in the decay phase of the flare, as the rise phase of the flare usually happens on a much shorter timescale, and hence fewer data points are available in which to search for a signal. QPPs have been detected in the rise phase of some solar flares \citep[e.g.][]{2003SoPh..218..183F}, however, and there is evidence of additional peaks in the rise phase of some of the flares observed by \emph{Kepler}, so this could be the subject of a further study. The SAP data were used rather than the pre-search data conditioning simple aperture photometry (PDCSAP) light curves, which have systematic artefacts removed, because some PDCSAP light curves were found to have artificial periodicities introduced. The target pixel files were also inspected to check that the flares were not due to contamination from a nearby object (within 4 arcseconds from the target, the size of 1 Kepler pixel), by ensuring that the pixels showing the flare coincided with the target point spread function. Our sample of flares includes those found by \citet{2015MNRAS.450..956B}, and those described as having complex structure by \citet{2014ApJ...797..122D} and \citet{2015EP&S...67...59M}, however not all flares identified by these authors were considered as QPP flares. Flares with only two peaks were omitted as it is difficult to determine whether two peaks counts as a true periodicity, and only flares exhibiting some kind of periodicity in the wavelet spectrum and autocorrelation function were included in our sample. In addition to this, it is not possible to know whether the two peaks are part of the same flare, or are due to separate flares in different active regions. On the other hand, it is far less likely that more than two flares occurred in separate active regions at around the same time.

On 2016 February 4, a problem with the \emph{Kepler} short-cadence data pixel calibration was reported, affecting around half of the targets and meaning that those affected may have signals introduced by other stars falling on the same CCD column. Fortunately, in most cases the amount of contamination is very small compared to the target signal. The long-cadence data are unaffected, so in order to assess the reliability of the data used in this paper the short-cadence and long-cadence target pixel files were compared, and in nearly all cases the contamination was not visible. The long-cadence and short-cadence light curves were also compared, to ensure that the flares studied appeared in both. The stars where a small amount of contamination was visible in the target pixel data are KIC 2852961 (although the QPPs can also be seen in the long-cadence data for this star), KIC 5475645, KIC 10206340 and KIC 11560431, so results from these stars should be treated with caution until the problem is fixed in next data release.

Performing a wavelet transform on a time series gives a map of non-stationary power as a function of time and period, where the period has a range from double the time step up to the total duration of the time series. Unlike a windowed Fourier transform, there is no temporal scale imposed by defining a window size. Instead, the wavelet function is scaled for each time step, so that the full range of possible periods can be mapped accurately. The disadvantage of this method is that the resolution is generally poor, and there is an intrinsic uncertainty relating to the choice of mother wavelet and number of oscillations present in the mother wavelet \citep{2004SoPh..222..203D}. In this study the Morlet wavelet with the default wavenumber of 6 was chosen, to give better period resolution, but lower time resolution. The wavelet transforms were used to identify periodicities above the 99\% confidence level in the flare decay phases, with a duration greater than the period. In order to estimate the QPP period from the wavelet plot, along with the associated uncertainty, the global wavelet spectrum (a time-average of the wavelet spectrum) was plotted, and a Gaussian line profile fitted to the peak corresponding to the periodicity in the data.

The autocorrelation function (the correlation of a sample with itself) as a function of time lag is defined as:
\begin{equation}
   P(l) = \frac{\displaystyle\sum_{i=0}^{N-l-1}(x_i - \bar{x})(x_{i+l} - \bar{x})}{\displaystyle\sum_{i=0}^{N-1}(x_i - \bar{x})^2} ,
\end{equation}
where $x_i(i=0,...,N-1)$ is a time series, $\bar{x}$ is the mean of the time series, and $l$ is the time lag. The autocorrelation function of a periodic signal is also periodic, but any noise is substantially suppressed, so it is useful for enhancing any stable periodicities in the data and determining the quality of the periodicities.

\subsection{Modelling the flares}
\label{sec:modflare} 

In order to enhance short-term variability, and hence make any periodic behaviour easier to identify in the wavelet spectra and autocorrelation functions, the decay trends needed to be removed from the flare decay light curves. The following expressions were used to fit the flare decay trends using a least-squares method:
\begin{equation}
   F(t) = A_0e^{-t/t_{0}} + Bt + C,
   \label{eq:flarelin}
\end{equation}
or,
\begin{equation}
   F(t) = A_0e^{-t/t_{0}} + B(t-D)^2 + C,
   \label{eq:flarequad}
\end{equation}
where $F$ is the flux, $t$ is time, and $A_0$, $t_{0}$, $B$, $C$, $D$ are constants. In some cases a simple exponential decay fits the flare decay very well, however most light curves have underlying trends, albeit with timescales much longer than that of the flare. These trends can be the result of differential velocity aberration, orbital motion due to the star being a binary, rotational variability due to starspots, and/or transits. To account for this, additional terms appear in the above expressions, where in Equation~(\ref{eq:flarelin}) a linear term is added and in Equation~(\ref{eq:flarequad}) a parabolic term is added. For the cases where there was no background trend, $B$ was set to zero. Previous research has found that many flares are better fitted with a two-phase exponential model rather than a single exponential decay \citep[e.g.][]{2014ApJ...797..122D}, since different regimes where conductive or radiative post-flare cooling can exist \citep[e.g.][]{1995ApJ...439.1034C}. The focus of this study is the QPPs, however, and since a simple single exponential model for the flare decay gave good fits, the impact of the inclusion of a second exponential component on the QPP parameters obtained would have been minor. Furthermore, the QPPs disrupt the smooth shape of the flare decay, making it difficult to accurately fit a two-component model. Hence a more simplistic model seemed more appropriate for this study. Whether Equation \ref{eq:flarelin} or Equation \ref{eq:flarequad} was used to fit the flare decay was decided based on the shape of the light curve in the vicinity of the flare. Both of these trend functions are aperiodic, and hence their subtraction from the original light curve cannot introduce artificial periodicities.

\begin{figure*}
	\includegraphics[width=0.85\linewidth]{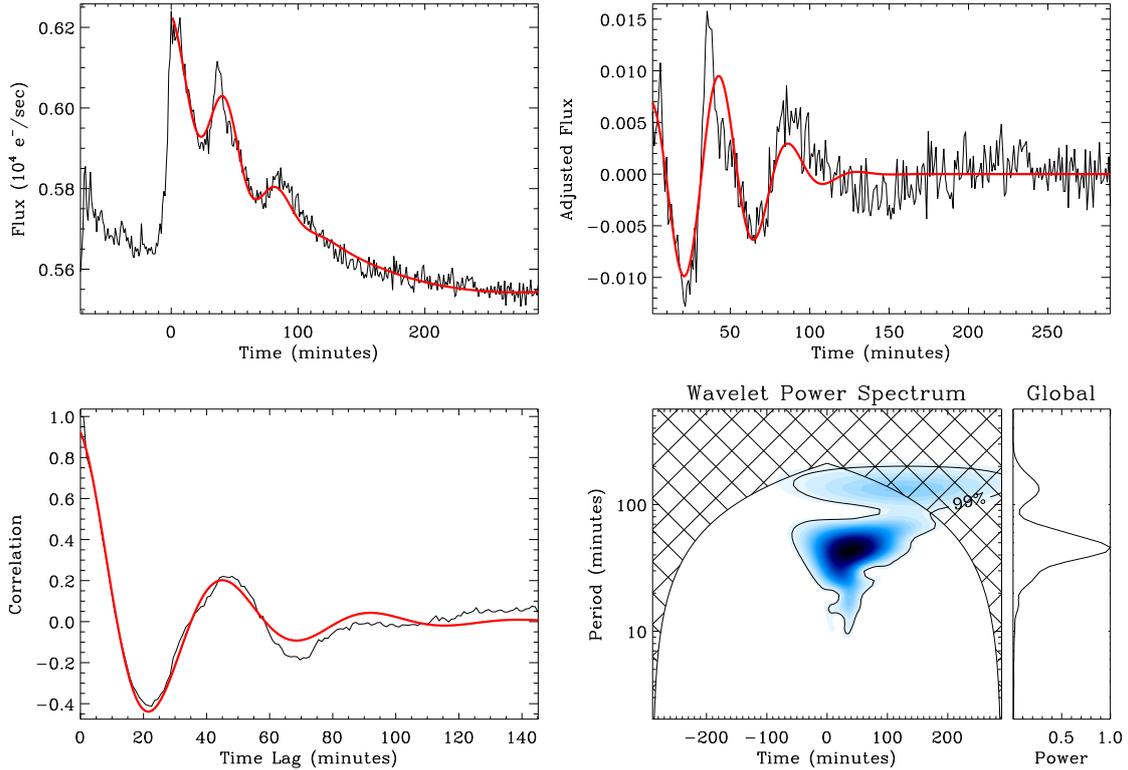}
    \caption{\emph{Top left:} light curve showing the decay phase of a flare on KIC 12156549, where start time is at the flare peak (MJD 55287.92 for this flare). The red overplotted line shows the result of a least-squares fit to the flare decay combined with the QPPs. \emph{Top right:} the same light curve as in Figure~\ref{fig:flare}, but with a fit to the flare decay trend subtracted in order to emphasise short-term variability. The red overplotted line shows a decaying sinusoidal fit. \emph{Bottom left:} The autocorrelation function of the time series shown in the top right panel, with a fitted exponentially decaying sinusoid shown in red. \emph{Bottom right:} the wavelet spectrum of the time series in the top right plot, which has been padded with zeros at the beginning in order to bring the feature of interest into the centre of the cone of influence. The spectrum shows a feature at a period of around 45\,minutes. The far-right panel shows the global wavelet spectrum.}
    \label{fig:flare}
\end{figure*}

Figure~\ref{fig:flare} shows stages of the analysis of a single flare on the star KIC 12156549. The decay phase light curve of this flare is shown in the top left panel, and the top right panel shows the same light curve after the flare decay trend has been subtracted. Performing a wavelet transform on the detrended light curve gives the contour plot in the bottom right panel, which shows a prominent feature above the 99\,\% confidence level at a period of around 45\,minutes, and suggests the presence of an oscillatory signal. Finally, the bottom left panel shows the autocorrelation of the detrended light curve, and more clearly shows a decaying sinusoidal signal. A fit to this plot gives a period of ($46 \pm 1$)\,minutes.

The majority of flares with a QPP-like signal in the light curve and wavelet plot either had a non-constant period, or did not decay in a straight-forward manner, meaning that the QPPs could not easily be fitted. Some of the flares, however, did have a QPP signal which appeared to be a stable decaying oscillation (i.e. a decaying signal which undergoes at least two cycles of oscillation and has a constant period) in their wavelet and autocorrelation plots. For these 11 flares (shown in Figures~\ref{fig:flare} and \ref{fig:flare1}-\ref{fig:flare10}) the QPPs were fitted simultaneously with the underlying decay trend of the flare, using one of the following expressions combined with either Equation~(\ref{eq:flarelin}) or (\ref{eq:flarequad}):
\begin{equation}
   \centering
   F(t) = A\exp\Bigg(\frac{-(t-B)}{\tau_{e}}\Bigg)\cos\bigg(\frac{2\pi}{P}t + \phi\bigg),
   \label{eq:qppexp}
\end{equation}
or,
\begin{equation}
   F(t)  = A\exp\Bigg(\frac{-(t-B)^2}{2\tau_{g}^2}\Bigg)\cos\bigg(\frac{2\pi}{P}t + \phi\bigg),
   \label{eq:qppgau}
\end{equation}
where $P$ is the period, $\phi$ is the phase, $\tau_e$ and $\tau_g$ are the exponential and Gaussian damping times, respectively, and $A$, $B$ are constants. An exponential decay is a natural assumption when considering damped harmonic oscillators, and has already been detected in stellar flares \citep[e.g.][]{2013ApJ...773..156A}, however MHD oscillations may also have a Gaussian damping profile. This was first discovered in numerical simulations by \citet{2012A&A...539A..37P}, then justified analytically by \citet{2013A&A...551A..39H}, and shown in observations by \citet{2016A&A...585L...6P}. The choice of which decay profile to use when fitting the flares was made based on the reduced chi-square goodness-of-fit test, although for two of the flares the Gaussian modulated fit (Equation~(\ref{eq:qppgau})) did not converge, hence the exponential decay model was chosen (Equation~(\ref{eq:qppexp})). These fits allowed a more precise estimation of the QPP period to be obtained, along with an estimation of the QPP decay time. Uncertainties for these fitted parameters are based on the uncertainties of \emph{Kepler} flux values, and were obtained by performing 10,000 Monte Carlo simulations. The error resampling method was used, where for each simulated sample a set of random numbers was drawn from the error distributions for each of the \emph{Kepler} SAP flux values. These 10,000 simulated samples were fitted using Equation~(\ref{eq:qppexp}) or (\ref{eq:qppgau}), where the initial parameters used for these fits were the same as those obtained by fitting the original light curve. Gaussian functions were then fitted to the resulting histograms of the fitted parameters, in order to obtain a robust estimate of the value and uncertainty of each parameter. The red overplotted line in Figure~\ref{fig:flare} shows the fit to the flare on KIC 12156549, which gave a QPP period of $44.6 \pm 0.6$\,minutes. Figures showing the analysis of the other flares with stable decaying oscillations are shown in Appendix~\ref{sec:B}.

\subsection{Calculating flare energies}
\label{sec:flareen}

The energies of the superflares were estimated using a similar method to \citet{2013ApJS..209....5S}. The flare luminosity ($L_{\mathrm{flare}}$) is a function of time, and integrating with respect to time gives the total energy radiated ($E_{\mathrm{flare}}$): 
\begin{equation}
   E_{\mathrm{flare}} = \int_{\mathrm{flare}} L_{\mathrm{flare}}(t)\,\mathrm{d}t.
   \label{eq:int}
\end{equation}
The bolometric flare luminosity can be calculated from the Stefan-Boltzmann relation, assuming that the emission can be approximated by a blackbody spectrum:
\begin{equation}
   L_{\mathrm{flare}}(t) = \sigma T^{4}_{\mathrm{flare}}A_{\mathrm{flare}}(t),
\end{equation}
where $\sigma$ is the Stefan-Boltzmann constant, $T_{\mathrm{flare}}$ is the effective temperature of the flare, and $A_{\mathrm{flare}}$ is the area covered by the flare. The ratio of the measured flare flux and stellar flux should be equal to the ratio of the flare luminosity and stellar luminosity, since $L = 4\pi d^2F$ (where $d$ is the distance to the star and $F$ is the measured flux), and the observed flare and stellar luminosities can be written as:
\begin{equation}
   L_{\mathrm{flare, obs}} = A_{\mathrm{flare}}(t)\int r_{\lambda}B_{\lambda(T_{\mathrm{flare}})}\,\mathrm{d}\lambda,
\end{equation}
and
\begin{equation}
   L_{\mathrm{\star, obs}} = \pi R^2_{\star}\int r_{\lambda}B_{\lambda(T_{\mathrm{eff}})}\,\mathrm{d}\lambda,
\end{equation}
where $\lambda$ is the wavelength, $r_{\lambda}$ is the \emph{Kepler} instrument response function, $B_{\lambda(T)}$ is the Planck function, and $R_{\star}$ is the stellar radius. Hence we can write: 
\begin{equation}
   \frac{\Delta F(t)}{F(t)} = \frac{A_{\mathrm{flare}}(t)\int r_{\lambda}B_{\lambda(T_{\mathrm{flare}})}\,\mathrm{d}\lambda}{\pi R_{\star}^2\int r_{\lambda}B_{\lambda(T_{\mathrm{eff}})}\,\mathrm{d}\lambda},
\end{equation}
where $\frac{\Delta F}{F}$ is the change in flux due to the flare, normalised by the underlying stellar flux. This expression can be used to find $A_{\mathrm{flare}}(t)$, which can then be used to find $L_{\mathrm{flare}}(t)$ and $E_{\mathrm{flare}}$. The underlying trend in the light curve is approximately linear in the vicinity of a flare, since the duration of the flare is typically short compared to the timescale of light curve modulation due to the rotation of the star. Hence, the flare amplitude, $\frac{\Delta F}{F}$, was found by subtracting the linear interpolation between the flare start and flare end from the measured flux, and then dividing by the same linear interpolation. Uncertainties of the flare energies were estimated by performing Monte Carlo simulations with the stellar temperature and radius uncertainties (given in Table~\ref{tab:starparams}), as well as the estimated uncertainty of the flare temperature, which was taken to be $9000 \pm 500$\,K \citep{1992ApJS...78..565H, 2003ApJ...597..535H, 2011A&A...530A..84K}. Other sources of uncertainty include assuming that the star and flare behave like blackbody radiators, defining the flare start and end (for example, the exponential decay nature of the flare decline makes it difficult to determine exactly when the flare has ended), the flux uncertainties, and the limited cadence of the data when performing the integration in Equation~(\ref{eq:int}). The latter three uncertainties will be much smaller than the others, hence have a negligible effect on the flare energy uncertainty.

\subsection{Stellar parameters}

\begin{table*}
   \caption{Parameters of the stars which have a flare showing evidence of QPPs. For each star the Kepler Input Catalogue (KIC) number is given, along with the temperature, radius, two estimates of the rotation period obtained using different methods, surface gravity and the Kepler magnitude ($K_p$). Temperatures, radii and surface gravities were taken from \citet{2014ApJS..211....2H}, with the exception of KIC 9726699, for which the temperature and radius are based on its M4 classification \citep{2004AJ....128..463R}. The note EB indicates that the star is an eclipsing binary, which prevented some rotation periods from being obtained without the use of a more complex modelling procedure. $^{(a)}$ For this star, the rotation period was obtained by \citet{2015ApJ...806..212D}.}
   \label{tab:starparams}
   \begin{tabular}{c c c c c c c c}
   \hline
   KIC	     & Temp.                 &Radius                      &Rot. per. from    &Rot. per. from               &log\,$g$                    &$K_{p}$   &Notes   \\	
   	     & (K)                   &(R$_{\odot}$)               &wavelet (days)    &autocorr. (days)             &(cm/s$^2)$                  &          &        \\	
   \hline	                                                          \\
   2852961   &$4882^{+126}_{-118}$   &$5.910^{+3.154}_{-2.027}$   &$35.5 \pm 4.8$    &$35.505 \pm 0.054$           &$2.888^{+0.355}_{-0.313}$   &10.146    &Subgiant \\[3pt]	
   3128488   &$4565^{+123}_{-127}$   &$0.546^{+0.037}_{-0.041}$   &$6.09 \pm 0.73$   &$6.171 \pm 0.014$            &$4.698^{+0.055}_{-0.032}$   &11.667    &        \\[3pt]	
   3540728   &$5015^{+173}_{-168}$   &$0.559^{+0.031}_{-0.034}$   &$2.10 \pm 0.25$   &$2.1472 \pm 0.0018$          &$4.697^{+0.048}_{-0.028}$   &12.596    &        \\[3pt]
   4671547   &$4175^{+160}_{-184}$   &$0.657^{+0.038}_{-0.057}$   &$8.03 \pm 0.96$   &$8.2215 \pm 0.0048$          &$4.649^{+0.039}_{-0.041}$   &11.293    &        \\[3pt]
   4758595   &$3573^{+88}_{-77}$     &$0.400^{+0.080}_{-0.050}$	  &$19.5 \pm 2.7$    &$19.51 \pm 0.16$             &$4.858^{+0.060}_{-0.100}$   &12.148    &        \\[3pt]
   5475645   &$5513^{+171}_{-141}$   &$0.708^{+0.193}_{-0.046}$   &$7.27 \pm 0.87$   &$7.504 \pm 0.012$	           &$4.627^{+0.029}_{-0.161}$   &11.205    &        \\[3pt]	
   6184894   &$5388^{+175}_{-142}$   &$0.717^{+0.172}_{-0.069}$   &$2.60 \pm 0.32$   &$2.640 \pm 0.014$            &$4.560^{+0.072}_{-0.216}$   &13.028    &        \\[3pt]	
   6437385   &$5727^{+180}_{-224}$   &$2.061^{+1.476}_{-1.270}$   &$13.4 \pm 1.6$    &$13.6154 \pm 0.0016$         &$3.707^{+0.757}_{-0.360}$   &11.539    &        \\[3pt]
   7664485   &$5510^{+166}_{-134}$   &$0.762^{+0.221}_{-0.054}$   &$3.12 \pm 0.38$   &$3.145 \pm 0.016$            &$4.594^{+0.029}_{-0.177}$   &13.264    &        \\[3pt]
   7885570   &$5587^{+165}_{-134}$   &$0.756^{+0.316}_{-0.053}$   &$0.85 \pm 0.10$   &$0.9139 \pm 0.0062$          &$4.593^{+0.031}_{-0.246}$   &11.679    &EB      \\[3pt] 
   7940533   &$5495^{+169}_{-133}$   &$0.798^{+0.320}_{-0.070}$	  &$3.82 \pm 0.46$   &$3.9032 \pm 0.0031$          &$4.543^{+0.045}_{-0.244}$   &12.862    &EB      \\[3pt] 
   8226464   &$6028^{+153}_{-160}$   &$1.535^{+0.707}_{-0.490}$   &$3.08 \pm 0.37$   &$3.130 \pm 0.014$            &$4.044^{+0.318}_{-0.228}$   &11.468    &        \\[3pt]
   8414845   &$5693^{+162}_{-134}$   &$0.899^{+0.388}_{-0.121}$   &$1.88 \pm 0.22$   &$1.8889 \pm 0.0072$          &$4.436^{+0.111}_{-0.275}$   &13.298    &        \\[3pt]
   8915957   &$5518^{+138}_{-123}$   &$2.652^{+2.406}_{-0.354}$   &$46.8 \pm 5.8$    &$46.40 \pm 0.58$             &$3.467^{+0.121}_{-0.429}$   &10.918    &Subgiant \\[3pt]
   9641031   &$6126^{+147}_{-166}$   &$1.176^{+0.459}_{-0.204}$   &--		     &--                           &$4.285^{+0.160}_{-0.236}$   &9.177     &EB      \\[3pt]
   9652680   &$5825^{+145}_{-146}$   &$0.835^{+0.314}_{-0.067}$   &$1.41 \pm 0.17$   &$1.430 \pm 0.014$            &$4.555^{+0.032}_{-0.267}$   &11.210    &        \\[3pt] 
   9655129   &$5334^{+173}_{-141}$   &$0.810^{+0.458}_{-0.095}$   &--		     &--  	                   &$4.492^{+0.091}_{-0.471}$   &13.805    &EB      \\[3pt]
   9726699   &$3100^{+150}_{-300}$   &$0.26^{+0.13}_{-0.06}$      &--                &$0.592596 \pm 0.00021^{a}$   &$5.283$                     &12.738    &        \\[3pt]
   9821078   &$4268^{+136}_{-140}$   &$0.680^{+0.030}_{-0.058}$   &--                &$9.792 \pm 0.015$            &$4.602^{+0.048}_{-0.017}$   &14.117    &EB      \\[3pt]
   9946017   &$6799^{+172}_{-220}$   &$2.892^{+0.885}_{-1.555}$   &$1.41 \pm 0.17$   &$1.430 \pm 0.014$            &$3.655^{+0.529}_{-0.182}$   &11.146    &        \\[3pt]
   10206340  &$5759^{+112}_{-120}$   &$0.945^{+0.183}_{-0.056}$	  &$2.25 \pm 0.28$   &$2.28150 \pm 0.00065$        &$4.481^{+0.039}_{-0.160}$   &11.203    &EB      \\[3pt] 
   10459987  &$5153^{+146}_{-135}$   &$0.649^{+0.095}_{-0.042}$   &$5.98 \pm 0.74$   &$6.048 \pm 0.023$            &$4.658^{+0.030}_{-0.084}$   &10.625    &        \\[3pt]
   10528093  &$5334^{+170}_{-140}$   &$0.746^{+0.190}_{-0.079}$   &$12.2 \pm 1.5$    &$12.1180 \pm 0.0089$         &$4.536^{+0.079}_{-0.270}$   &13.563    &        \\[3pt] 
   11551430  &$5648^{+108}_{-91}$    &$1.605^{+0.377}_{-0.345}$   &$4.10 \pm 0.52$   &$4.1652 \pm 0.0036$          &$4.019^{+0.183}_{-0.132}$   &10.691    &        \\[3pt]
   11560431  &$5367^{+223}_{-175}$   &$0.828^{+0.322}_{-0.082}$   &$3.06 \pm 0.37$   &$3.1609 \pm 0.0052$          &$4.514^{+0.060}_{-0.245}$   &9.694     &        \\[3pt] 
   11560447  &$5105^{+147}_{-152}$   &$0.593^{+0.042}_{-0.038}$   &--                &$0.4891 \pm 0.0020$          &$4.665^{+0.050}_{-0.037}$   &10.834    &EB      \\[3pt] 
   11610797  &$6140^{+140}_{-193}$   &$1.048^{+0.591}_{-0.090}$   &$1.58 \pm 0.19$   &$1.6303 \pm 0.0026$          &$4.455^{+0.039}_{-0.348}$   &11.535    &        \\[3pt] 
   11665620  &$4683^{+132}_{-132}$   &$0.573^{+0.043}_{-0.045}$   &$0.358 \pm 0.042$ &$0.32693 \pm 0.00038$        &$4.676^{+0.058}_{-0.034}$   &14.242    &        \\[3pt]
   12102573  &$4474^{+161}_{-137}$   &$0.745^{+0.029}_{-0.057}$	  &$2.71 \pm 0.32$   &$2.74038 \pm 0.00065$        &$4.561^{+0.046}_{-0.022}$   &11.815    &        \\[3pt]
   12156549  &$5888^{+350}_{-356}$   &$1.043^{+0.464}_{-0.191}$   &$3.61 \pm 0.42$   &$6.653 \pm 0.014$            &$4.373^{+0.137}_{-0.280}$   &15.886    &        \\[3pt]
   \hline
   \end{tabular}
\end{table*}

The surface temperature, radius, rotation period, surface gravity and Kepler magnitude for each of the stars with evidence of QPPs in one or more flares is given in Table~\ref{tab:starparams}. The stellar temperatures, radii and surface gravities are taken from \citet{2014ApJS..211....2H}, and the Kepler magnitudes are taken from the Kepler Input Catalogue. Stellar rotation periods were obtained following the method described in \citet{2016MNRAS.455.3110A}, which we summarise here. We use the PDCSAP detrended \emph{Kepler} data for this step \citep{Stumpe:2012bj, Smith:2012ji}, and note that this can attenuate signals arising from rotation periods over approximately 21 days \citep{Garcia:2013td}. Rotation periods are determined using both the autocorrelation function and a wavelet analysis, the latter to ensure the period arises from the entire duration of the data rather than a single isolated region in time. These are calculated as described in Section~\ref{sec:id}. Rotation periods are extracted from the autocorrelation function by first smoothing it with a Gaussian filter with standard deviation the same as the strongest detected peak, truncated at $3.1\sigma$. The first 4 harmonics of this peak are then identified, and a linear fit performed to the peak periods, as well as the maximum peak and the origin. The gradient and error of this fit then gives our autocorrelation function period and its error. We then confirm this extracted period visually against the \emph{Kepler} light curve. This follows the similar method proposed and tested in \citet{McQuillan:2014gp}. While the wavelet analysis, due to its nature, gives less precise period measurements, we extract the wavelet period for comparison, as performed in, for example, \citet{Garcia:2014ds, Mathur:2014cz}. The wavelet power is summed over the time axis, giving the global wavelet spectrum. This is fit by the sum of several Gaussian profiles, with the given period found from the largest amplitude peak and its error from the half width at half maximum of this peak.

\section{RESULTS AND DISCUSSION}
\label{sec:results}

\begin{figure*}
	\includegraphics[width=\linewidth]{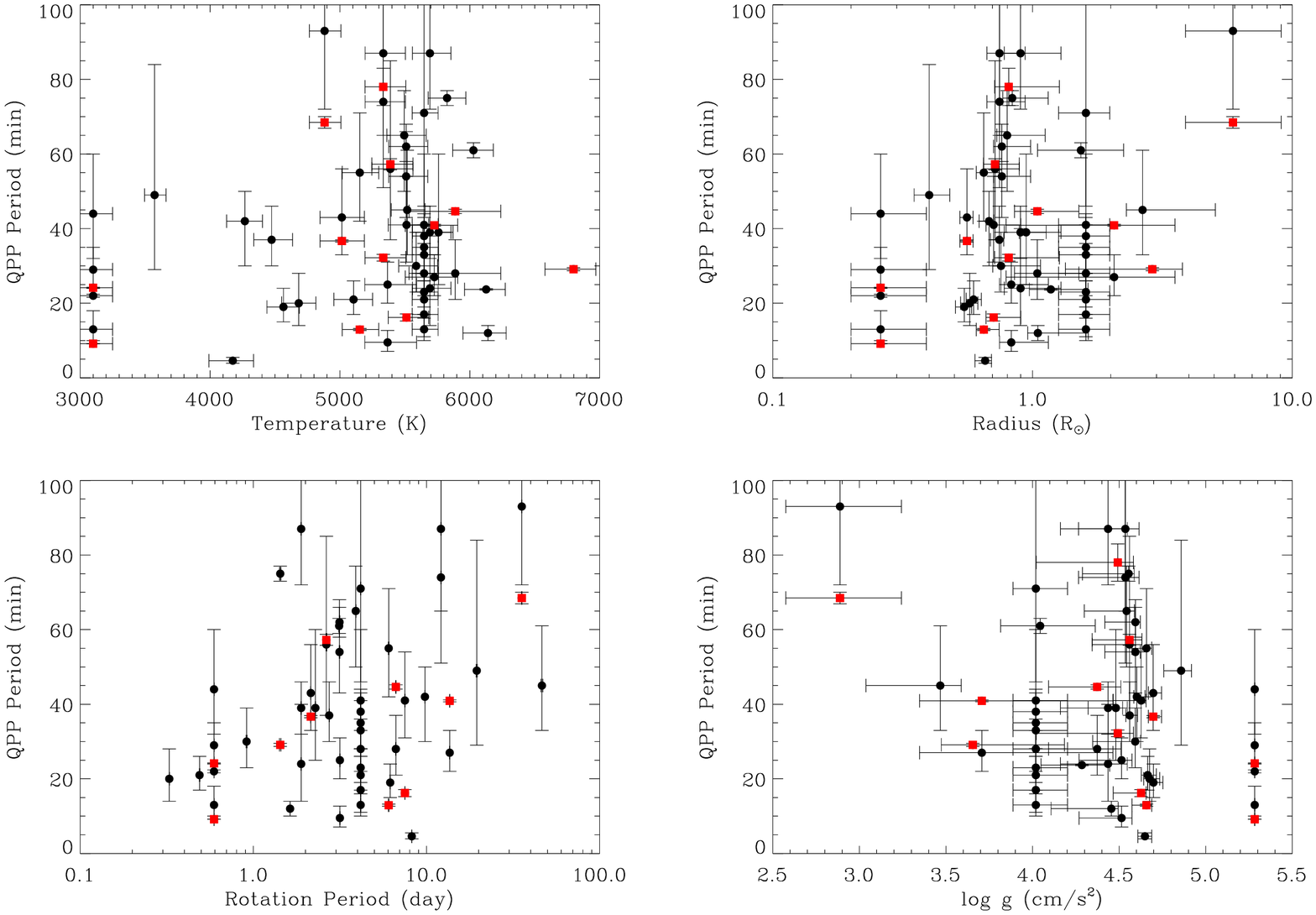}
	\caption{\emph{Top left:} scatter plot of stellar effective temperature and QPP period. The Pearson correlation coefficient is 0.184, with a p-value of 0.086, and the Spearman coefficient is 0.077, with a p-value of 0.567. \emph{Top right:} scatter plot of stellar radius and QPP period. The Pearson correlation coefficient is 0.312, with a p-value of 0.009, and the Spearman coefficient is 0.172, with a p-value of 0.200. \emph{Bottom left:} scatter plot of stellar rotation period and QPP period. The Pearson correlation coefficient is 0.357, with a p-value of 0.003, and the Spearman coefficient is 0.246, with a p-value of 0.073. \emph{Bottom right:} scatter plot of stellar surface gravity and QPP period. The Pearson correlation coefficient is -0.281, with a p-value of 0.017, and the Spearman coefficient is -0.193, with a p-value of 0.151. In all plots, the red square points indicate the flares with a stable decaying oscillation, and the black round points indicate the quasi-periodic flares.}
	\label{fig:pervsstparams}
\end{figure*}

\begin{table*}
\centering
\caption{Parameters of the flares showing evidence of a QPPs. The KIC number of the star is given along with the approximate time at which the flare occurs (given as the Modified Julian Date, equivalent to the Julian Date with 2400000.5 days subtracted), the QPP period, and an estimate of the energy released during the flare.}
\label{tab:flareparams1}
\begin{tabular}{ c c c c c c c c c }
\cline{1-4}\cline{6-9}
KIC        &Time (Modified  &Period              &Flare energy                            &   &KIC        &Time (Modified  &Period              &Flare energy \\
	   &Julian Date)    &(min)               &(erg)                                   &   &	          &Julian Date)    &(min)               &(erg)        \\
\cline{1-4}\cline{6-9}	\\                  
2852961	   &55238.22        &$68 \pm 2$          &$(7.5^{+10.3}_{-4.3}) \times 10^{35}$   &   &9726699    &55749.56        &$29^{+6}_{-5}$      &$(1.6^{+2.1}_{0.9}) \times 10^{32}$     \\[3pt]
2852961    &55240.27        &$93^{+27}_{-21}$    &$(1.4^{+2.0}_{-0.8}) \times 10^{37}$    &   &9726699    &55999.77        &$9.16 \pm 0.02$     &$(2.6^{+3.8}_{1.5}) \times 10^{33}$     \\[3pt]
3128488    &54990.32        &$19^{+5}_{-4}$      &$(1.6^{+0.5}_{-0.4}) \times 10^{34}$    &   &9726699    &56082.84        &$44^{+16}_{-12}$    &$(1.6^{+2.2}_{-0.9} \times 10^{32}$     \\[3pt]
3540728    &55751.38        &$43^{+13}_{-10}$    &$(2.4^{+0.8}_{-0.6}) \times 10^{34}$    &   &9821078    &55487.25        &$42^{+8}_{-12}$     &$(6.0^{+2.2}_{-1.6}) \times 10^{33}$    \\[3pt]
3540728    &55807.25        &$36.7 \pm 0.3$      &$(4.1^{+1.4}_{-1.0}) \times 10^{34}$    &   &9946017    &55217.57        &$29.1 \pm 0.4$      &$(6.9^{+12.2}_{-4.4}) \times 10^{35}$   \\[3pt]
4671547    &55090.04        &$4.6^{+0.9}_{-0.7}$ &$(1.8^{+0.8}_{-0.5}) \times 10^{33}$    &   &10206340   &55076.74        &$39^{+21}_{-14}$    &$(2.5^{+1.0}_{-0.7}) \times 10^{34}$    \\[3pt]
4758595    &56219.15        &$49^{+35}_{-20}$    &$(1.6^{+0.9}_{-0.6}) \times 10^{33}$    &   &10459987   &55158.15        &$12.9 \pm 0.3$      &$(2.5^{+0.9}_{-0.7}) \times 10^{33}$    \\[3pt]
5475645    &55095.92        &$16.2 \pm 0.9$      &$(1.3^{+0.6}_{-0.4}) \times 10^{34}$    &   &10459987   &56189.39        &$55^{+16}_{-13}$    &$(1.8^{+0.6}_{-0.5}) \times 10^{33}$    \\[3pt]
5475645    &56330.43        &$41^{+13}_{-10}$    &$(1.8^{+0.9}_{-0.6}) \times 10^{34}$    &   &10528093   &56214.53        &$87^{+29}_{-22}$    &$(2.4^{+1.3}_{-0.8}) \times 10^{34}$    \\[3pt]
6184894    &56243.87        &$57 \pm 1$          &$(2.2^{+1.1}_{-0.7}) \times 10^{34}$    &   &10528093   &56262.77        &$74^{+34}_{-23}$    &$(1.4^{+0.8}_{-0.5}) \times 10^{35}$    \\[3pt]
6184894    &56291.77        &$56^{+29}_{-19}$    &$(3.5^{+1.8}_{-1.2}) \times 10^{34}$    &   &11551430   &55004.60        &$21^{+7}_{-5}$      &$(1.4^{+1.0}_{-0.6}) \times 10^{35}$    \\[3pt]
6437385    &55391.10        &$27^{+6}_{-5}$      &$(5.1^{+17.9}_{-4.0}) \times 10^{35}$   &   &11551430   &55024.13        &$28^{+8}_{-6}$      &$(1.3^{+0.8}_{-0.5}) \times 10^{35}$    \\[3pt]
6437385    &55393.76        &$40.9 \pm 0.3$      &$(6.1^{+20.2}_{-4.7}) \times 10^{35}$   &   &11551430   &55031.05        &$38 \pm 5$          &$(3.4^{+2.2}_{-1.3}) \times 10^{35}$    \\[3pt]
7664485    &56107.70        &$54^{+14}_{-11}$    &$(3.3^{+1.6}_{-1.1}) \times 10^{34}$    &   &11551430   &55031.96        &$35^{+9}_{-7}$      &$(1.7^{+1.1}_{-0.7}) \times 10^{35}$    \\[3pt]
7664485    &56119.79        &$62 \pm 4$          &$(2.7^{+1.4}_{-0.9}) \times 10^{34}$    &   &11551430   &56117.13        &$33^{+8}_{-7}$      &$(3.3^{+2.1}_{-1.3}) \times 10^{35}$    \\[3pt]
7885570    &55010.88        &$30^{+9}_{-7}$      &$(2.4^{+1.4}_{-0.9}) \times 10^{34}$    &   &11551430   &56134.74        &$23^{+5}_{-4}$      &$(2.6^{+1.7}_{-1.0}) \times 10^{35}$    \\[3pt]
7940533    &55317.42        &$65^{+12}_{-15}$    &$(7.9^{+5.2}_{-3.1}) \times 10^{34}$    &   &11551430   &56166.63        &$71^{+40}_{-25}$    &$(2.3^{+1.5}_{-0.9}) \times 10^{35}$    \\[3pt]
8226464    &55012.10        &$61 \pm 2$          &$(2.2^{+2.7}_{-1.2}) \times 10^{35}$    &   &11551430   &56208.35        &$41^{+19}_{-13}$    &$(4.3^{+2.9}_{-1.7}) \times 10^{35}$    \\[3pt]
8414845    &56217.91        &$39 \pm 7$          &$(3.2^{+2.4}_{-1.4}) \times 10^{34}$    &   &11551430   &56264.09        &$13^{+4}_{-3}$      &$(4.5^{+3.0}_{-1.8}) \times 10^{35}$    \\[3pt]
8414845    &56285.43        &$87^{+18}_{-15}$    &$(5.7^{+4.3}_{-2.5}) \times 10^{34}$    &   &11551430   &56270.75        &$17^{+9}_{-6}$      &$(4.7^{+3.1}_{-1.9}) \times 10^{35}$    \\[3pt]
8414845    &56293.70        &$24^{+16}_{-10}$    &$(2.5^{+1.8}_{-1.0}) \times 10^{34}$    &   &11560431   &56150.68        &$9.5^{+3.2}_{-2.4}$ &$(2.1^{+1.5}_{-0.9}) \times 10^{33}$    \\[3pt]
8915957    &55152.31        &$45^{+16}_{-12}$    &$(8.4^{+11.4}_{4.8}) \times 10^{34}$    &   &11560431   &56193.12        &$25^{+6}_{-5}$      &$(3.1^{+2.1}_{-1.3}) \times 10^{33}$    \\[3pt]
9641031    &55614.55        &$23.7 \pm 0.2$      &$(6.9^{+5.6}_{-3.1}) \times 10^{34}$    &   &11560447   &55947.44        &$21^{+5}_{-4}$      &$(5.8^{+2.0}_{-1.5}) \times 10^{34}$    \\[3pt]
9652680    &55085.13        &$75 \pm 2$          &$(6.0^{+3.6}_{-2.3}) \times 10^{34}$    &   &11610797   &54981.63        &$12 \pm 2$          &$(1.1^{+0.9}_{-0.5}) \times 10^{35}$    \\[3pt]
9655129	   &56149.04        &$78 \pm 5$          &$(3.4^{+3.0}_{-1.6}) \times 10^{34}$    &   &11665620   &55762.95        &$20^{+8}_{-6}$      &$(4.2^{+1.4}_{-1.0}) \times 10^{34}$    \\[3pt]
9726699	   &55382.78        &$22.0 \pm 0.4$      &$(2.9^{+3.7}_{-1.6}) \times 10^{32}$    &   &12102573   &55086.03        &$37^{+9}_{-7}$      &$(3.7^{+1.3}_{-0.9}) \times 10^{33}$    \\[3pt]
9726699	   &55401.16        &$24.2 \pm 0.1$      &$(5.5^{+7.9}_{-3.2}) \times 10^{31}$    &   &12156549   &55287.92        &$44.6 \pm 0.6$      &$(5.0^{+4.7}_{-2.4}) \times 10^{35}$    \\[3pt]
9726699    &55409.48        &$13^{+5}_{-3}$      &$(7.8^{+11.2}_{-4.6}) \times 10^{31}$   &   &12156549   &55347.20        &$28^{+9}_{-7}$      &$(5.4^{+5.3}_{-2.7}) \times 10^{35}$    \\[3pt]
\cline{1-4}\cline{6-9}
\end{tabular}
\end{table*}

\begin{figure*}
	\includegraphics[width=\linewidth]{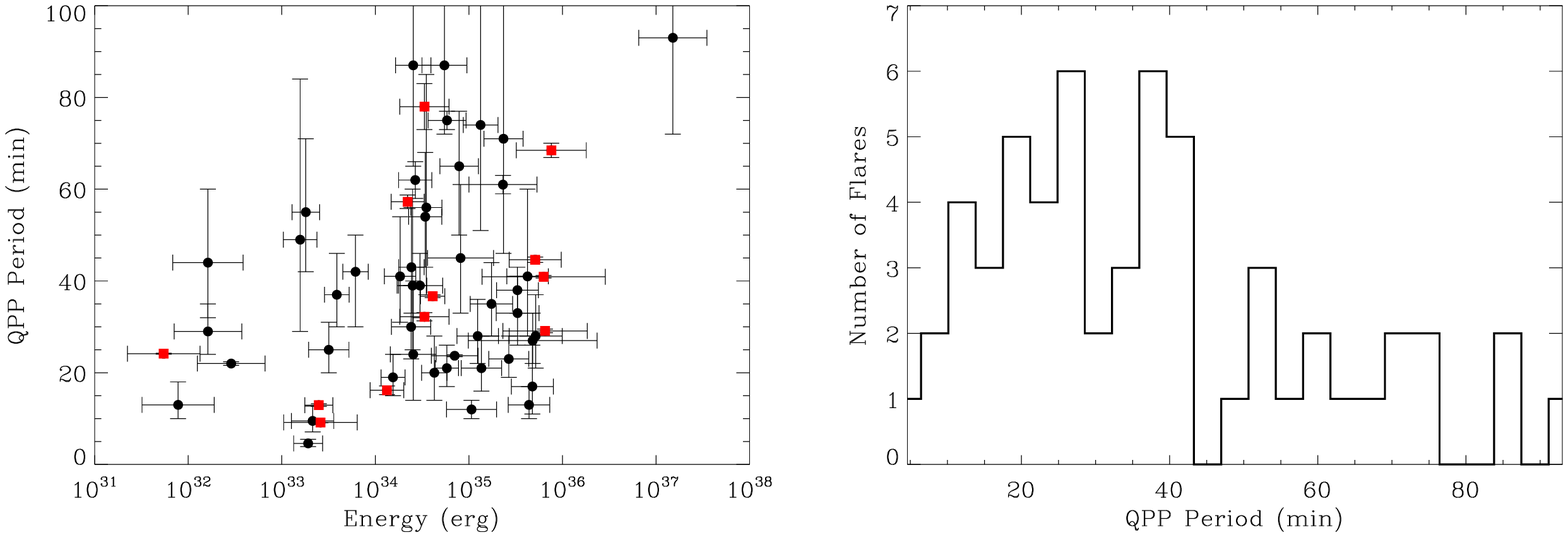}
    \caption{\emph{Left:} scatter plot of flare energy and QPP period. The Pearson correlation coefficient is 0.333, with a p-value of 0.005, and the Spearman coefficient is 0.219, with a p-value of 0.102. The red points indicate the flares with a high-quality, stable decaying oscillation. \emph{Right:} histogram showing the distribution of QPP periods.}
    \label{fig:pervsene_hist}
\end{figure*}

Out of the 1439 flares detected on 216 different stars, 56 are found to have pronounced QPP-like signatures in the light curve, of which 11 show evidence of stable decaying oscillations. The host stars range from F- to M-type. We note that 1439 is a lower estimate for the total number of flares, since low amplitude flares are not detected by the automated search. All available short cadence light curves for the flaring stars were checked by eye, however, to ensure that no flares with a QPP-like signature were missed. Figure~\ref{fig:pervsstparams} shows scatter plots of the period of the 56 flares showing a QPP-like signal with various stellar parameters: namely surface temperature, radius, rotation period and surface gravity. The surface gravity is dependent on the temperature and radius, but is included for completeness. Details of the flares used are given in Table~\ref{tab:flareparams1}, where the periods of the 11 flares with stable oscillations were obtained using the method described in Section~\ref{sec:modflare}, and for the other flares the period was estimated using the global wavelet spectrum. A complete list of parameters obtained from the flare fits are given in Table~\ref{tab:flarefitparams}. None of the correlation coefficients suggest a relationship between the QPP period and the global stellar parameters, implying that the pulsations are related to a local, rather than global, phenomenon. Nor is there a significant correlation between the QPP period and total flare energy, as shown in the left-hand panel of Figure~\ref{fig:pervsene_hist}. Since the flare energy scales with stellar temperature, radius, and surface gravity, as shown in Appendix~\ref{sec:A}, it follows that if the QPP period does not relate to any stellar parameter, then it should not relate to the flare energy either. In all of these plots, no distinction can be made between the flares with high-quality, stable decaying oscillations, and those which are quasi-periodic. The QPP decay times were also checked for any dependency on the stellar parameters or flare energy, but no significant correlations were found.

A histogram of the distribution of periods is given in the right-hand panel of Figure~\ref{fig:pervsene_hist}. While the detectable range of periods is limited by the cadence of the data and the duration of the flare, the plot shows that apart from the majority of flares having a period less than 45\,minutes, there does not appear to be a clear preference for a particular period range. Even for the cases where several QPP flares are detected on the same star, there is still a wide range of periods, as shown in Table~\ref{tab:flareparams1}. This is consistent with the solar case, where a wide range of QPP periods are detected in solar flares \citep[e.g.][]{2009SSRv..149..119N}. Little work has been done, however, to determine any relationships between the QPP period of solar flares and parameters of the flaring active region, so this could be the subject of future research.

\begin{figure*}
	\includegraphics[width=\linewidth]{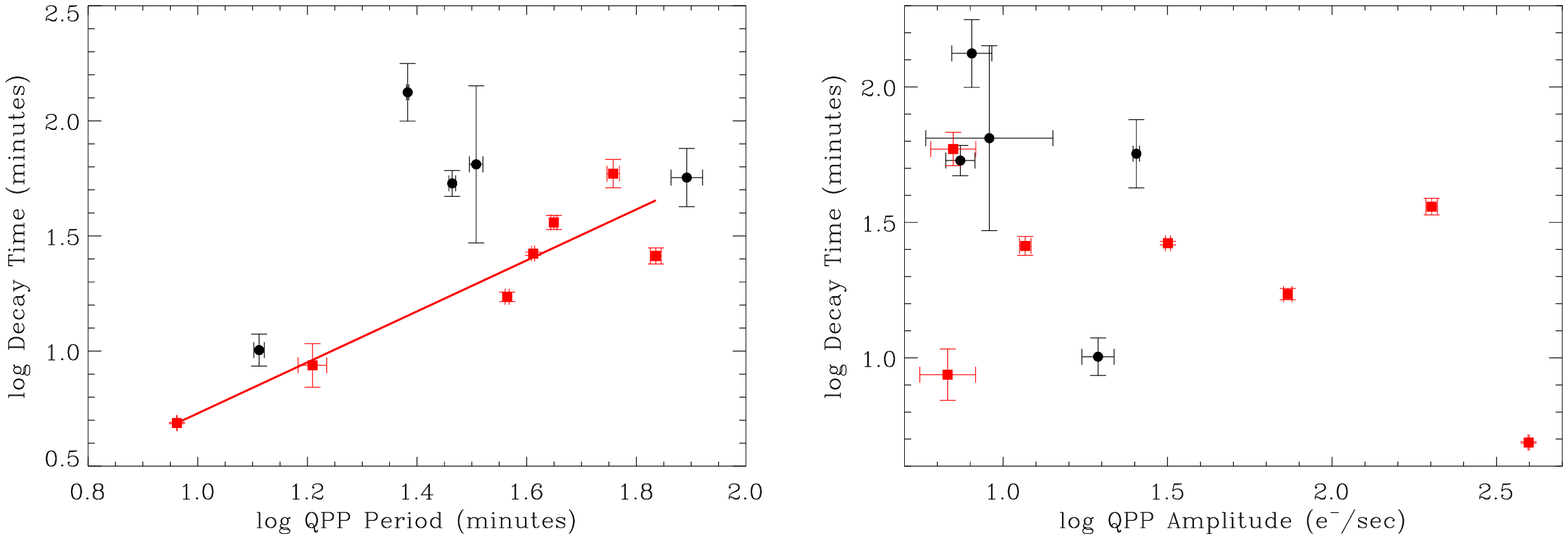}
    \caption{\emph{Left:} scatter plot of QPP period and QPP decay time, where only flares with a stable periodicity and an exponential or Gaussian decay profile were used. Those with an exponential decay profile are shown in black, and those with a Gaussian profile are shown in red. For the exponential decay profile flares, the Pearson correlation coefficient is 0.042, with a p-value of 0.378, and the Spearman coefficient is 0.300, with a p-value of 0.624. For the Gaussian modulated flares, the Pearson correlation coefficient is 0.737, with a p-value of 0.000, and the Spearman coefficient is 0.786, with a p-value of 0.036. \emph{Right:} scatter plot of normalised QPP amplitude and QPP decay time. The Pearson correlation coefficient is -0.371, with a p-value of 0.002, and the Spearman coefficient is -0.364, with a p-value of 0.245.}
    \label{fig:t0}
\end{figure*}

Properties of the flares with stable decaying oscillations are given in Table~\ref{tab:flareparams2}, with a complete list of parameters given in Table~\ref{tab:qppfitparams}. A scatter plot of the oscillation period and the oscillation decay time for these flares is given in the left-hand panel of Figure~\ref{fig:t0}, where the flares with exponentially damped oscillations do not show a significant correlation, but those with Gaussian modulated oscillations do. Due to the limited sample size, however, future observations will be needed to confirm this. Fitting a linear model for the flares with Gaussian modulated oscillations, as shown by the red line in the left-hand panel of Figure~\ref{fig:t0}, gives the following expression:
\begin{equation}
   \log \tau_{g} = (1.31 \pm 0.06)\log P - (0.46 \pm 0.07).
\end{equation}
The above expression is close to the relationship $\tau \sim P$, which has been derived both theoretically and observationally for coronal oscillations in the Sun, and corresponds to damping due to resonant absorption (also referred to as mode coupling) for kink waves \citep[e.g.][]{2002ApJ...576L.153O, 2003ApJ...598.1375A, 2016A&A...585A.137G} or thermal conduction for slow magnetoacoustic (longitudinal) waves \citep{2002ApJ...580L..85O}. The same damping mechanism can be responsible for the two types of damping profile; for example, for kink modes a low density contrast between the oscillating coronal loop and the background plasma results in weak mode coupling, and hence a Gaussian damping profile of the oscillations, while a high density contrast would result in strong mode coupling and hence exponential damping \citep{2016A&A...585L...6P}.  It is still possible that the exponentially damped oscillations could show a similar relationship with more data, since the decay time is not expected to scale perfectly with the period; both the decay time and period will also depend on other properties of the oscillating coronal structure.

The right-hand panel of Figure~\ref{fig:t0} shows that the decay times are not significantly correlated with the amplitude of the QPPs (normalised by the stellar flux at the base of the flare, so that the brightness of the star does not influence the amplitude) suggesting that the majority of these flares have a linear QPP signal, since a non-linear signal would result in higher amplitude QPPs being damped more strongly. There was also no evidence that the QPP period correlated with the amplitude.

\begin{table*}
\caption{Parameters of the flares showing evidence of a stable decaying oscillation. The KIC number of the star is given along with the approximate time at which the flare occurs, the period of oscillation, the decay time, the damping profile type that best fits the oscillations, and an estimate of the energy released during the flare. Note that KIC 9655129 is included twice as two separate periodicities were detected \citep{2015ApJ...813L...5P}. We also note that while the period obtained for the flare on KIC 9946017 is very close to the \emph{Kepler} long cadence time period of 29.4\,minutes, the same periodicity was not detected elsewhere in the light curve.}
\label{tab:flareparams2}
\begin{tabular}{c c c c c c c}
\hline
KIC        &Time (Modified  &Period from        &Period from       &Oscillation      &Decay       &Flare energy  \\
	   &Julian date)    &wavelet (min)      &fit (min)         &decay time (min) &profile     &(erg)         \\
\hline	\\                  
2852961	   &55238.22        &$60^{+16}_{-13}$   &$68 \pm 2$        &$26 \pm 2$       &Gaussian &$8.3 \times 10^{35}$ \\[3pt] 
3540728    &55807.25        &$36^{+10}_{-8}$    &$36.7 \pm 0.3$    &$17.2 \pm 0.8$   &Gaussian    &$4.2 \times 10^{34}$ \\[3pt]
5475645    &55095.92        &$16^{+5}_{-4}$     &$16.2 \pm 0.9$    &$9 \pm 2$        &Gaussian &$1.3 \times 10^{34}$ \\[3pt]
6184894    &56243.87        &$52^{+16}_{-12}$   &$57 \pm 1$        &$59 \pm 8$       &Gaussian    &$2.3 \times 10^{34}$ \\[3pt]
6437385    &55393.76        &$39^{+13}_{-10}$   &$40.9 \pm 0.3$    &$26.5 \pm 0.4$   &Gaussian    &$6.8 \times 10^{35}$ \\[3pt]
9655129	   &56149.04        &$78^{+23}_{-17}$   &$78 \pm 5$        &$58 \pm 15$      &Exponential &$3.6 \times 10^{34}$ \\[3pt]
9655129    &56149.04        &$32 \pm 7$         &$32 \pm 1$        &$65 \pm 37$      &Exponential &$3.6 \times 10^{34}$ \\[3pt]
9726699	   &55401.16        &$28^{+9}_{-7}$     &$24.2 \pm 0.1$    &$133 \pm 33$     &Exponential    &$5.5 \times 10^{31}$ \\[3pt]
9726699	   &55999.77        &$8 \pm 2$          &$9.16 \pm 0.02$   &$4.87 \pm 0.03$  &Gaussian &$2.6 \times 10^{33}$ \\[3pt]
9946017	   &55217.57        &$30^{+8}_{-6}$     &$29.1 \pm 0.4$    &$53 \pm 6$       &Exponential &$7.1 \times 10^{35}$ \\[3pt]
10459987   &55158.15        &$13^{+3}_{-2}$     &$12.9 \pm 0.3$    &$10 \pm 1$       &Exponential &$1.8 \times 10^{33}$ \\[3pt]
12156549   &55287.92        &$45^{+14}_{-11}$   &$44.6 \pm 0.6$    &$36 \pm 2$       &Gaussian    &$5.4 \times 10^{35}$ \\[3pt]
\hline
\end{tabular}
\end{table*}

\section{CONCLUSIONS}
\label{sec:conc}

Correlations between QPP periods in stellar flares and various parameters of the host stars were studied. The periods were not found to depend on any global stellar parameters, suggesting that they are related to local properties of the flaring site or active region only, rather than being the result of the leakage of global oscillations such as p- or g-modes (the periods of which are determined by stellar parameters). This supports the idea that the QPPs observed in stellar flares are akin to those in solar flares, and that coronal seismology techniques can be applied in the stellar context. The period was also found to be independent of the flare energy, again suggesting that QPPs in stellar superflares can be used to learn about the local conditions in flaring active regions, in much the same way as QPPs in solar flares. Another observational finding that supports the possible association of QPPs with MHD oscillations is the presence of a characteristic decay of the detected QPP signal in many cases, which is consistent with solar flare QPP light curves and spatially resolved oscillations of coronal plasma structures, observed in the extreme ultraviolet and microwave bands. This apparent independence of the QPP parameters from other observables makes them a potentially important independent diagnostic tool.

As expected from the theory of coronal loop oscillations, the QPP damping time was related to the period by a power law. Only the flares with Gaussian modulated QPPs showed a statistically significant correlation with a small p-value (where a p-value less than 0.05 indicates that the null hypothesis, i.e. that there is no correlation, should be rejected). On the other hand, the flares with exponentially damped QPPs showed no clear correlation, and the calculated p-value indicated that the correlation coefficient was not statistically significant, although we observe that the damping time does increase with the period overall. A larger sample size is necessary to confirm this finding, which hopefully can be achieved by future analysis of flares from the current K2 and \emph{XMM-Newton} missions.

It should be noted that the flares studied in this paper are all observed in white-light, and the origin of white-light emission in solar flares is currently not well understood. The QPP mechanisms discussed in the Introduction apply to microwave and X-ray emission via the gyrosynchrotron mechanism and bremsstrahlung, respectively, or thermal emission in the extreme ultraviolet and soft X-ray wavebands. Multiple wavelength observations of solar flares have, however, found that the white-light emission tends to be strongly associated with hard X-ray emission \citep{2003A&A...409.1107M, 2007ApJ...656.1187F}, and so it is expected that the same QPP mechanisms apply. The modulation of the non-thermal emission by MHD oscillations can be produced by either the modulation of the magnetic reconnection rate \citep{2006SoPh..238..313C, 2006A&A...452..343N}, or the kinematics of the non-thermal electrons \citep{1982SvAL....8..132Z}. An alternative interpretation could be magnetic dripping (an oscillatory regime of magnetic reconnection \citep[e.g.][]{2000A&A...360..715K}), but it does not explain the detected exponential and Gaussian damping scenarios of the oscillations.

\section*{ACKNOWLEDGEMENTS}

C.E.P. would like to thank Mark Hollands for useful discussions. We thank the anonymous referee for the helpful constructive comments. C.E.P. \& V.M.N.: This work was supported by the European Research Council under the \textit{SeismoSun} Research Project No. 321141. D.J.A. acknowledges funding from the European Union Seventh Framework programme (FP7/2007--2013) under grand agreement No. 313014 (ETAEARTH). V.M.N. acknowledges support from the STFC consolidated grant ST/L000733/1. A.-M.B. thanks the Institute of Advanced Study, University of Warwick for their support. We would like to thank the \emph{Kepler} team for providing the data used in this paper, obtained from the Mikulski Archive for Space Telescopes (MAST). Funding for the \emph{Kepler} mission is provided by the NASA Science Mission directorate. Wavelet software was provided by C. Torrence and G. Compo and is available at \\ \url{http://paos.colorado.edu/research/wavelets/}.




\bibliographystyle{mnras}
\bibliography{keplerqpp} 



\appendix

\section{Flare energy scaling laws}
\label{sec:A}

\begin{figure*}
	\includegraphics[width=\linewidth]{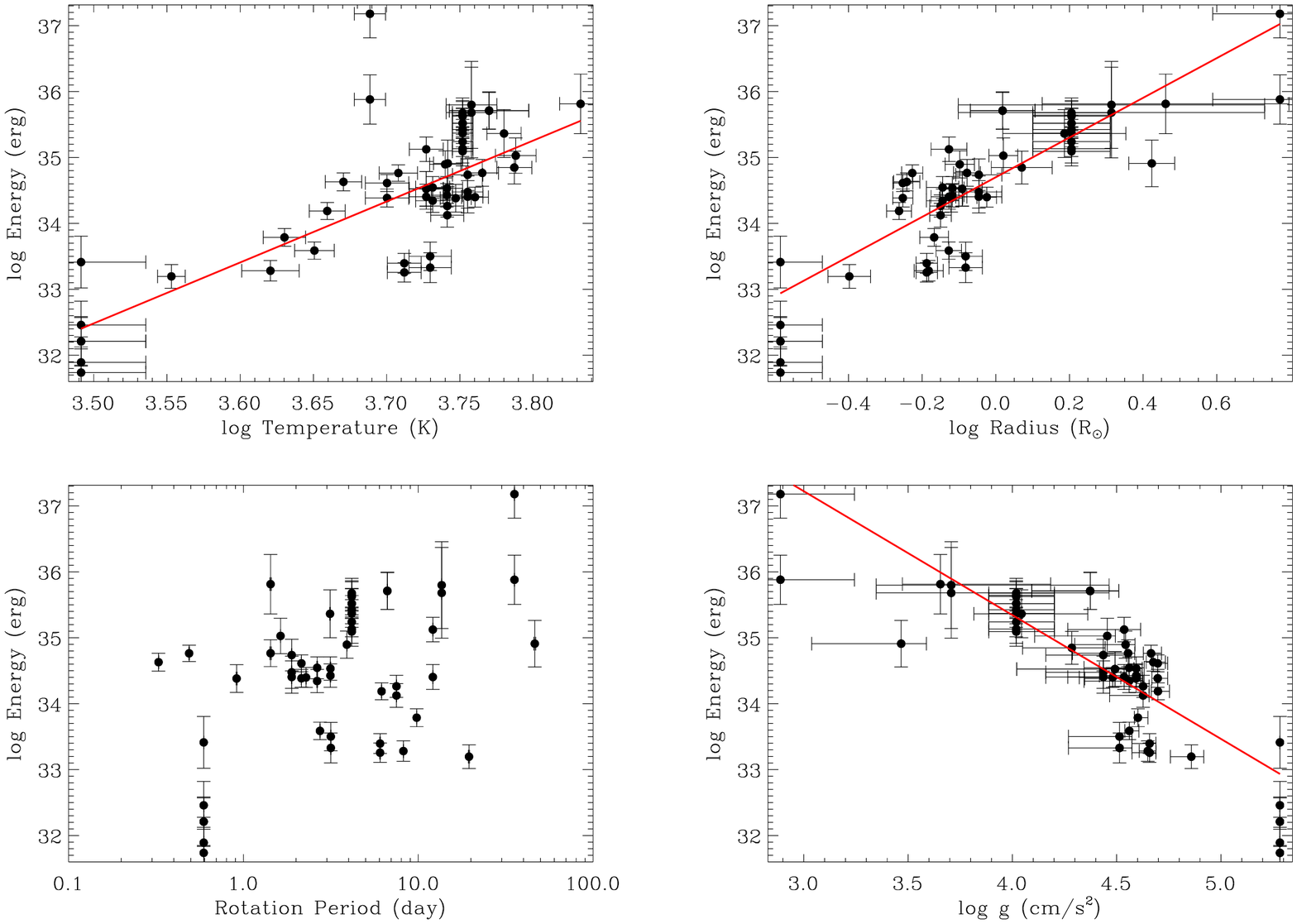}
	\caption{\emph{Top left:} scatter plot of stellar temperature and flare energy. The Pearson correlation coefficient is -0.001, with a p-value of 0.496, and the Spearman coefficient is 0.669, with a p-value of $10^{-8}$. \emph{Top right:} scatter plot of stellar radius and flare energy. The Pearson correlation coefficient is 0.651, with a p-value of $6 \times 10^{-9}$, and the Spearman coefficient is 0.858, with a p-value of $10^{-17}$. \emph{Bottom left:} scatter plot of stellar rotation period and flare energy. The Pearson correlation coefficient is 0.472, with a p-value of $8 \times 10^{-5}$, and the Spearman coefficient is 0.360, with a p-value of 0.007. \emph{Bottom right:} scatter plot of stellar surface gravity and flare energy. The Pearson correlation coefficient is -0.468, with a p-value of $10^{-4}$, and the Spearman coefficient is -0.858, with a p-value of $10^{-17}$.}
	\label{fig:enevsstparams}
\end{figure*}

The correlations of QPP parameters with stellar parameters and flare energies were studied in Section~\ref{sec:results}, and so in this section the dependency of the total flare energy on stellar parameters is examined. Plotting the flare energy against stellar surface temperature (as shown in the top left panel of Figure~\ref{fig:enevsstparams}) shows a strong positive correlation. While the Pearson correlation coefficient does not seem to be reliable for this case, due to the outlying points, the Spearman correlation coefficient and associated p-value shows a significant correlation, supporting previous findings, and fitting a straight line gives the following expression:
\begin{equation}
   \log E = (9.2 \pm 0.4)\log T_{\star} + (0 \pm 2).
\end{equation}
A similar relationship is found between the flare energy and stellar radius (top right panel of Figure~\ref{fig:enevsstparams}), with a highly statistically significant correlation, and the fitted expression is in excellent agreement with the relationship of $\log E = 3\log R_{\star}/R_{\sun} + 34.14$ found by \citet{2015MNRAS.447.2714B}, despite the smaller sample size used:
\begin{equation}
   \log E = (3.0 \pm 0.1)\log R_{\star}/R_{\sun} + (34.70 \pm 0.03).
\end{equation}
No significant correlation was found between the flare energy and stellar rotation period, as shown in the bottom left panel of Figure~\ref{fig:enevsstparams}, which is in agreement with previous findings \citep{2012Natur.485..478M, 2014ApJ...792...67C}. The bottom right panel of Figure~\ref{fig:enevsstparams} shows that the flare energy correlates negatively with stellar surface gravity, which is to be expected if larger, hotter main sequence stars tend to have a lower surface gravity. Fitting a linear expression gives:
\begin{equation}
   \log E = (-1.88 \pm 0.08)\log g + (42.9 \pm 0.4).
\end{equation}

Since no correlations were found between the QPP period and any stellar parameters or the flare energy, the relationships derived above are unlikely to have any bearing on the future study of QPPs in stellar flares. They may, however, have implications for the study of superflares: in particular, the likelihood of a superflare occuring on the Sun \citep{2013PASJ...65...49S}.

\section{Additional figures for flares with stable decaying oscillations}
\label{sec:B}

In this section plots are included for the other flares with stable decaying oscillations, as summarised in Table~\ref{tab:flareparams2}. The plots are equivalent to those described in Section~\ref{sec:modflare} for the star KIC 12156549.

\begin{figure*}
	\includegraphics[width=0.85\linewidth]{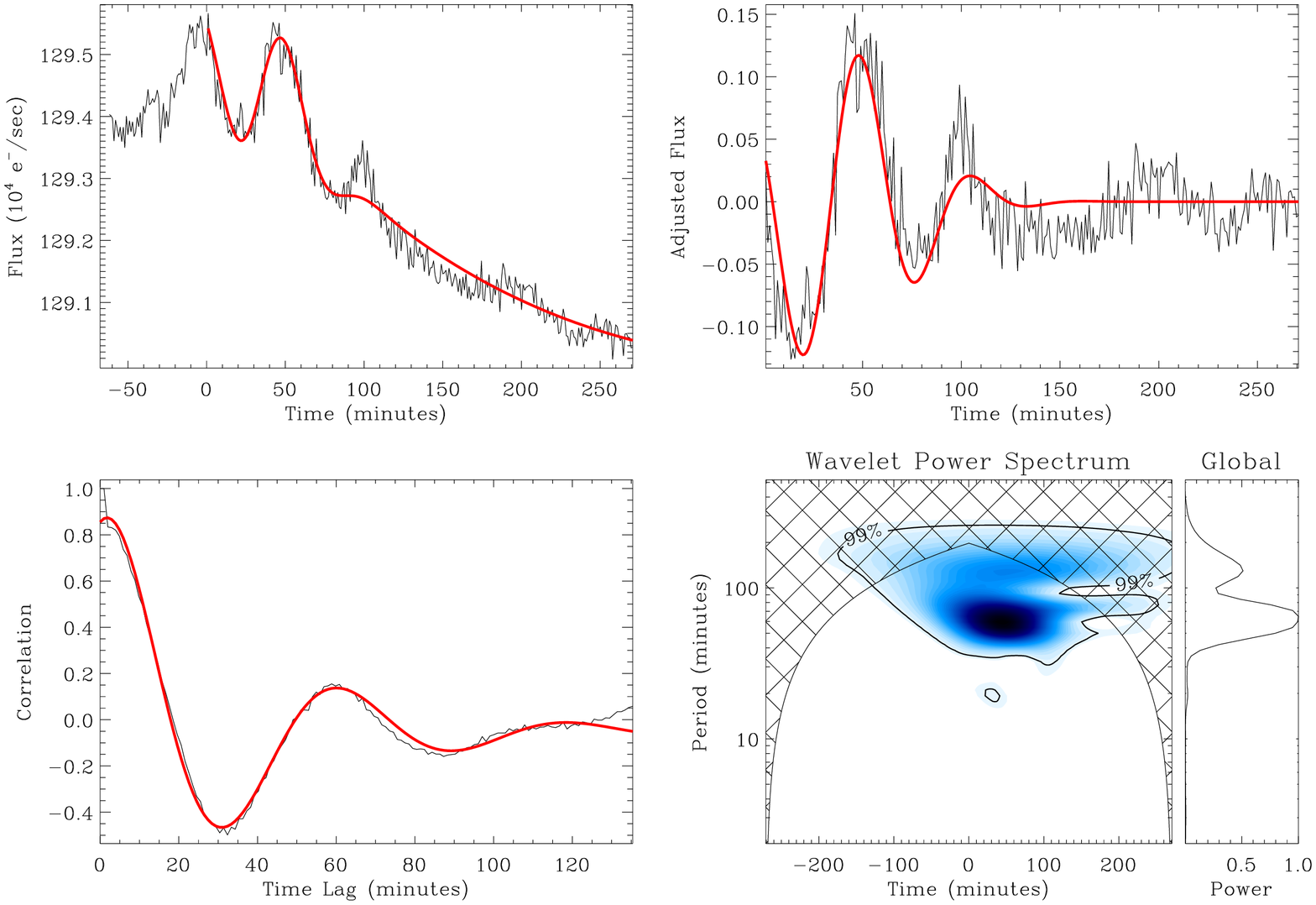}
    \caption{KIC 2852961, start time (MJD): 55238.22.}
    \label{fig:flare1}
\end{figure*}
\begin{figure*}
	\includegraphics[width=0.85\linewidth]{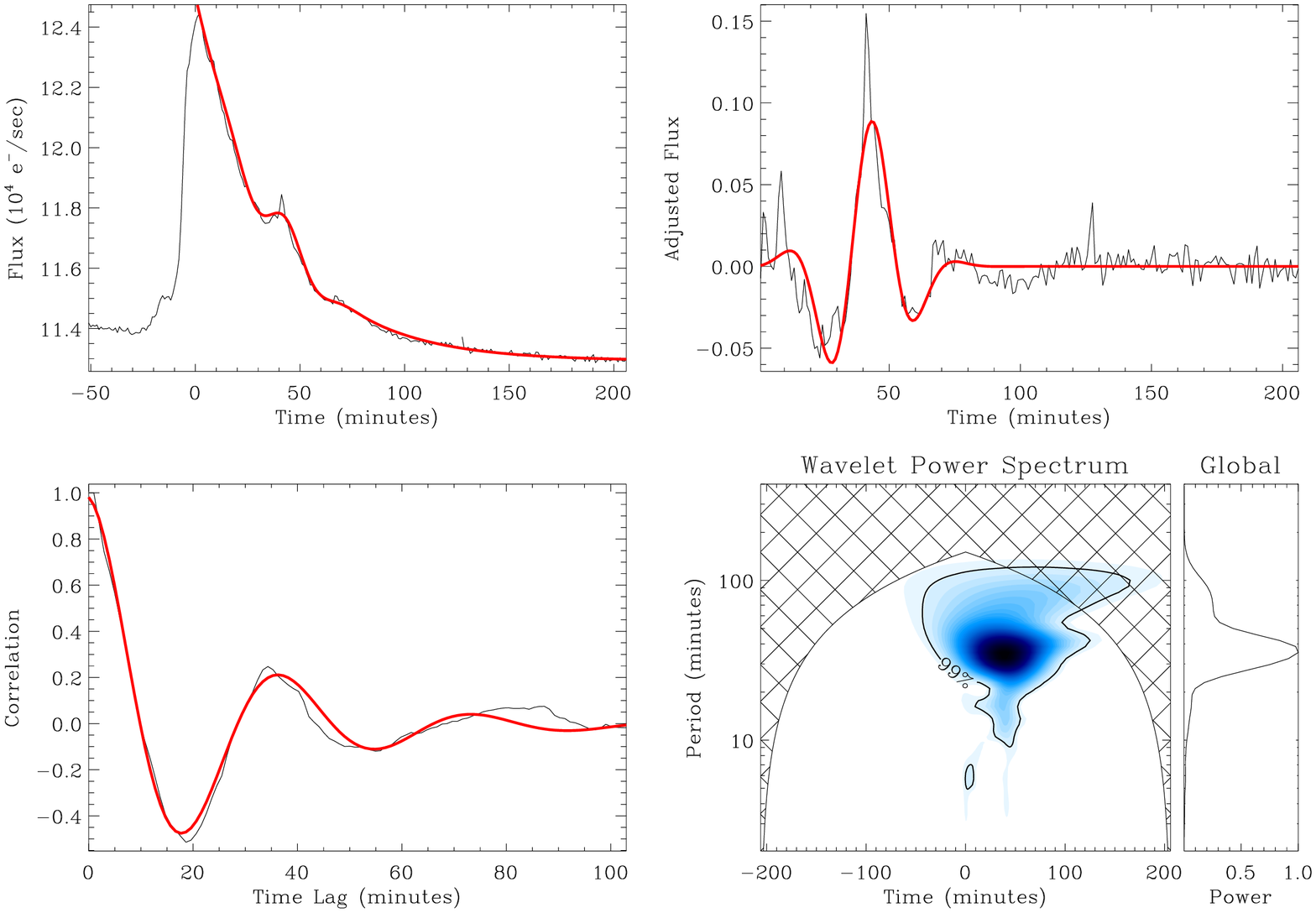}
    \caption{KIC 3540728, start time (MJD): 55807.25.}
    \label{fig:flare2}
\end{figure*}
\begin{figure*}
	\includegraphics[width=0.85\linewidth]{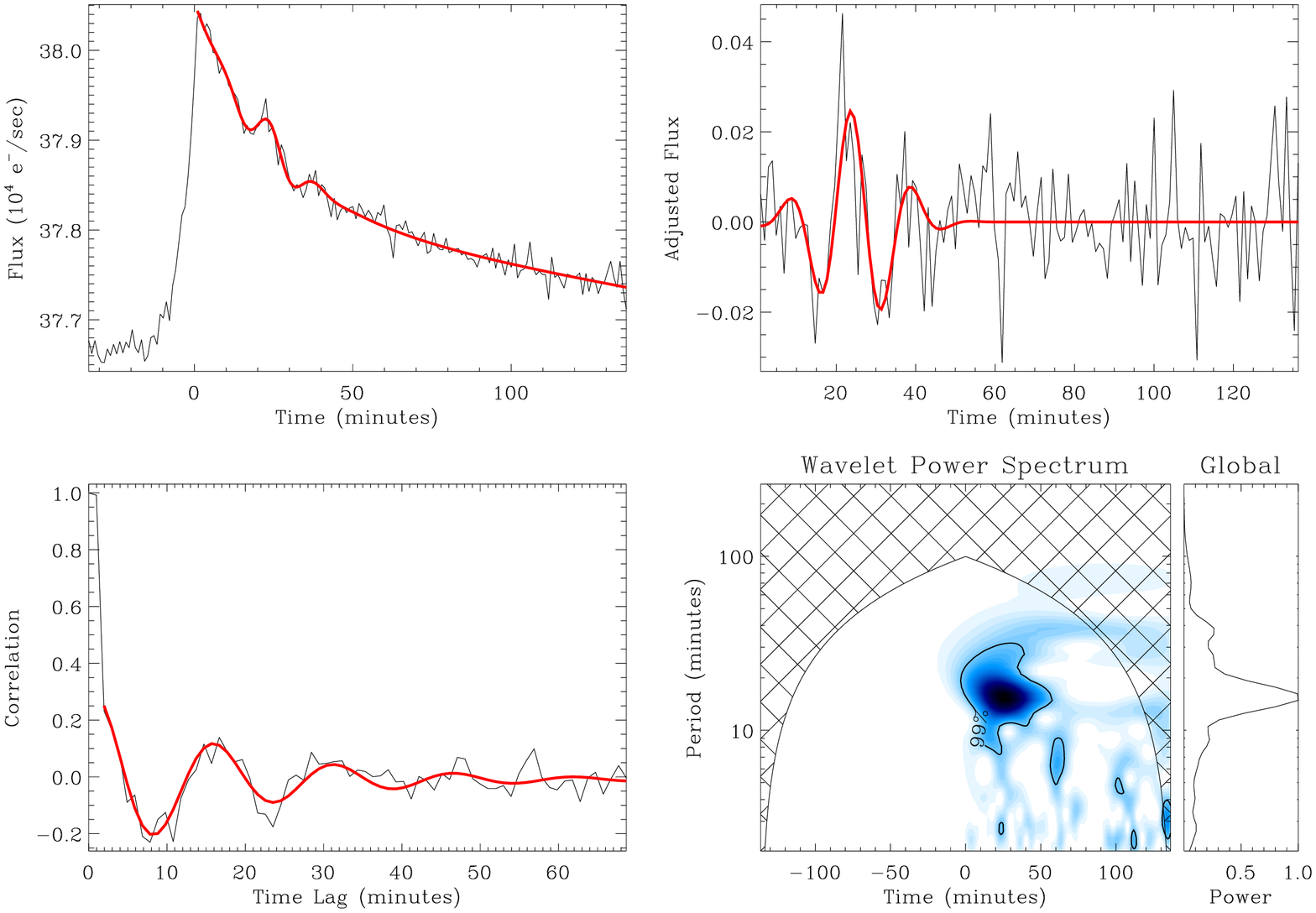}
    \caption{KIC 5475645, start time (MJD): 55095.92.}
    \label{fig:flare3}
\end{figure*}
\begin{figure*}
	\includegraphics[width=0.85\linewidth]{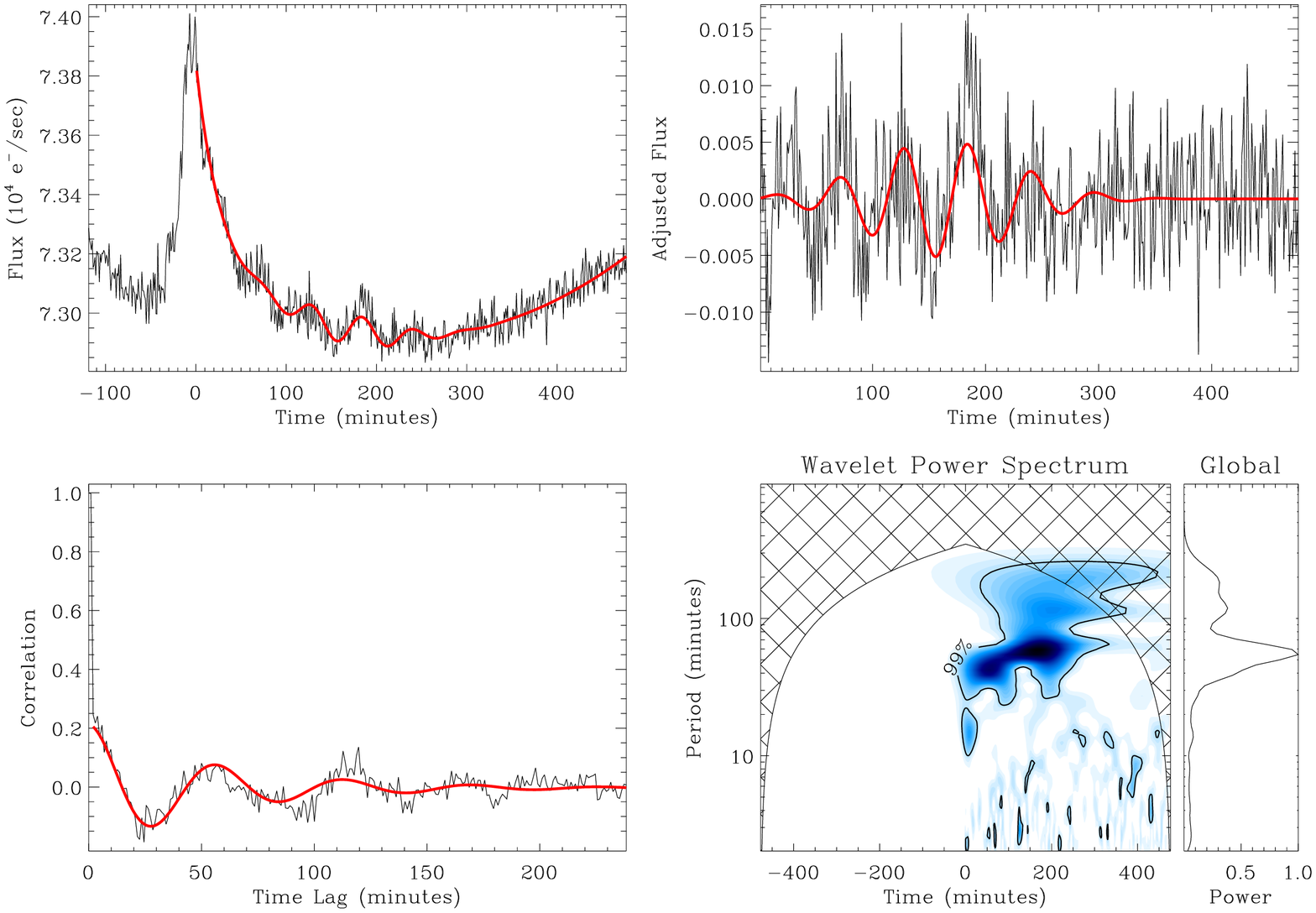}
    \caption{KIC 6184894, start time (MJD): 56243.87. The substantial underlying trend in the light curve is due to starspot modulation.}
    \label{fig:flare4}
\end{figure*}
\begin{figure*}
	\includegraphics[width=0.85\linewidth]{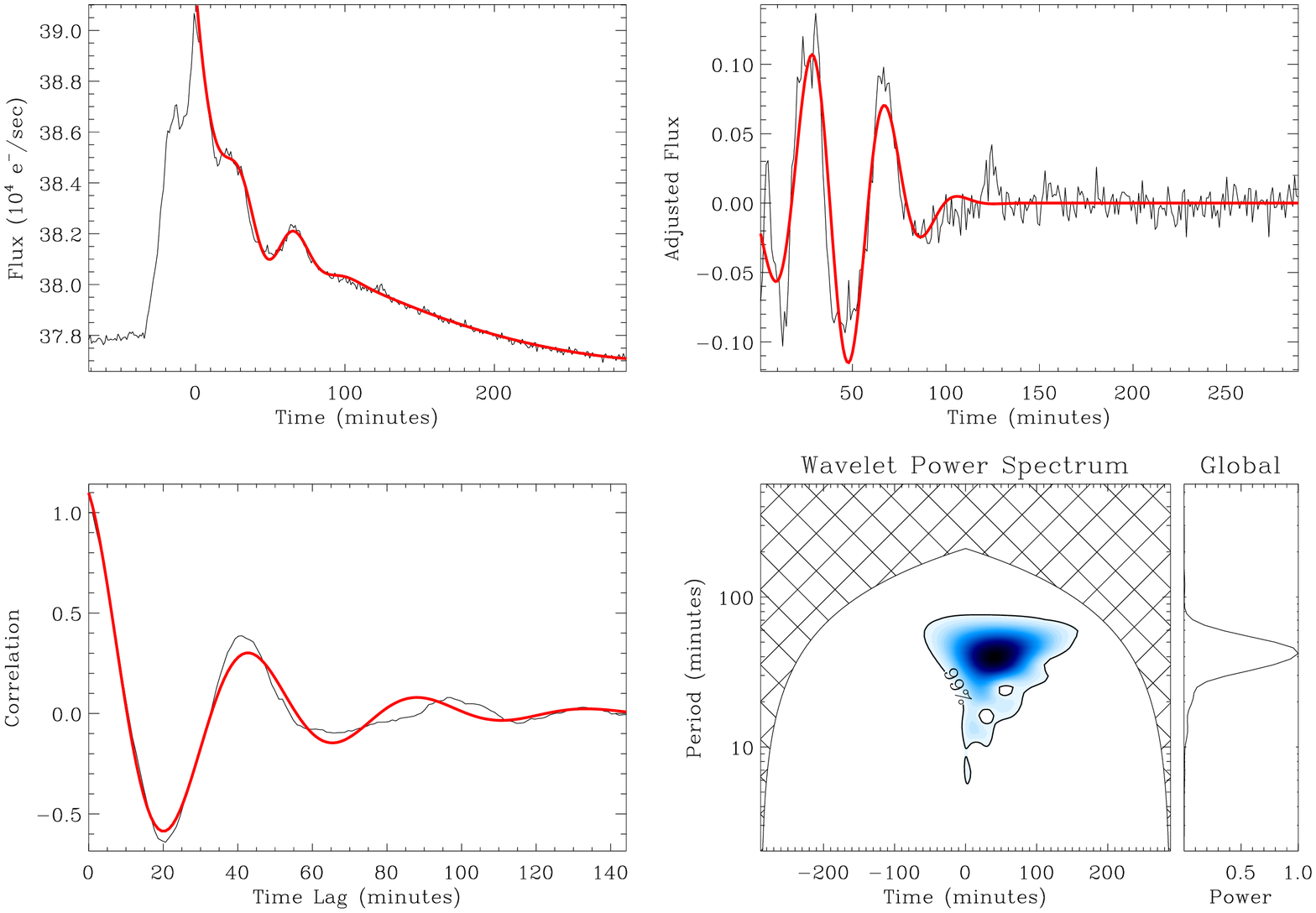}
    \caption{KIC 6437385, start time (MJD): 55393.76.}
    \label{fig:flare5}
\end{figure*}
\begin{figure*}
	\includegraphics[width=0.85\linewidth]{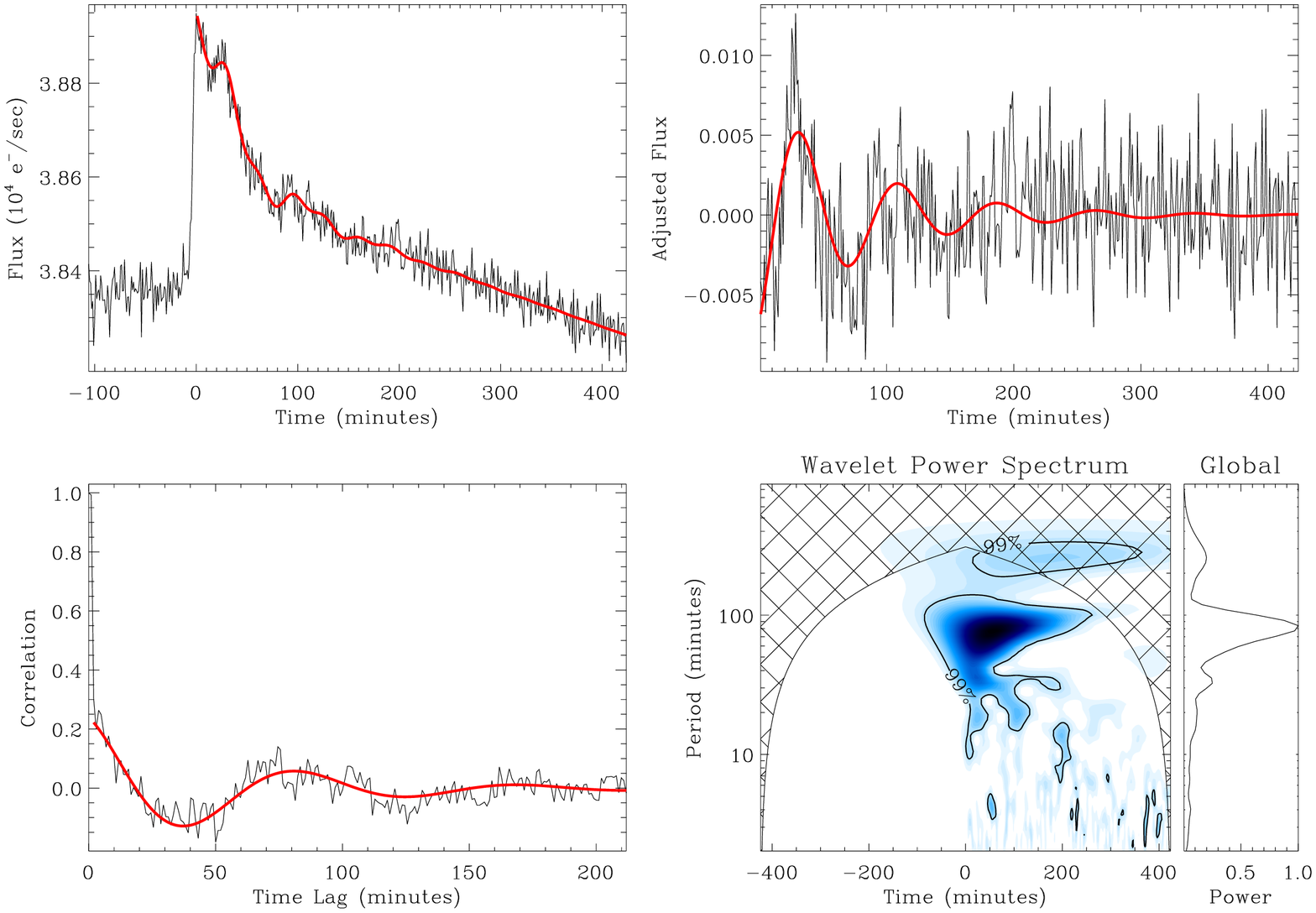}
    \caption{KIC 9655129, start time (MJD): 56149.04.}
    \label{fig:flare6}
\end{figure*}
\begin{figure*}
	\includegraphics[width=0.85\linewidth]{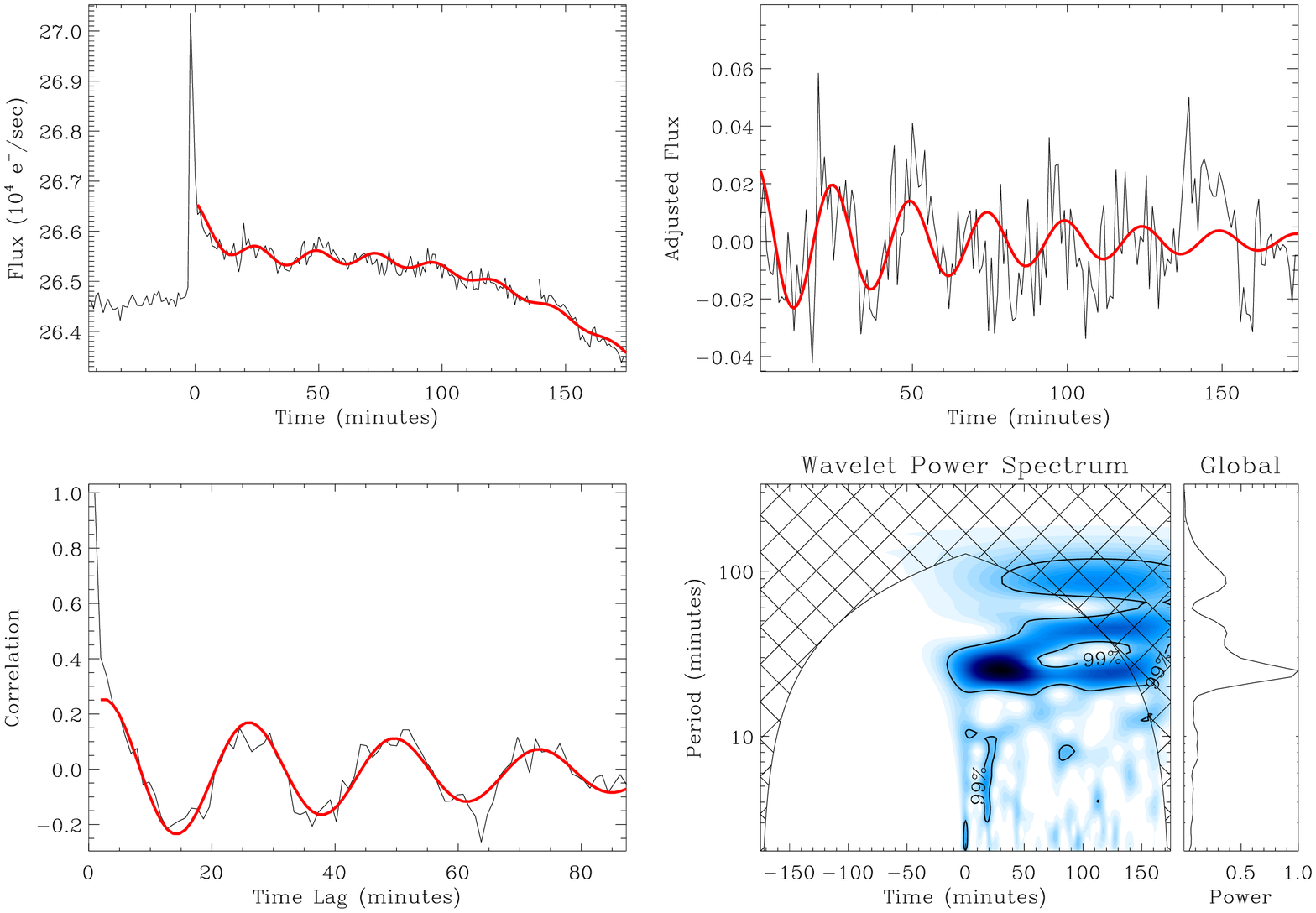}
    \caption{KIC 9726699, start time (MJD): 55401.16.}
    \label{fig:flare8}
\end{figure*}
\begin{figure*}
	\includegraphics[width=0.85\linewidth]{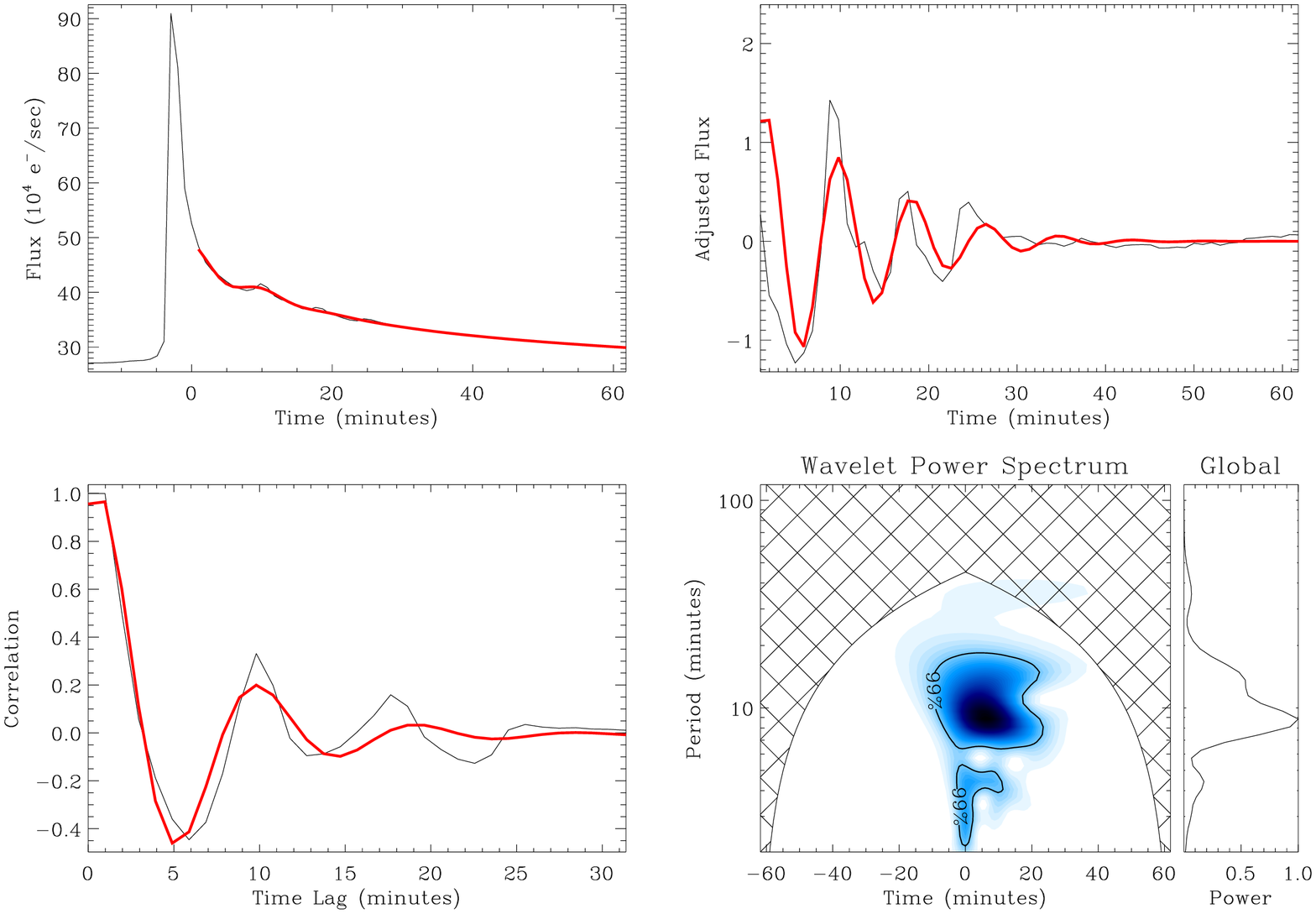}
    \caption{KIC 9726699, start time (MJD): 55999.77.}
    \label{fig:flare9}
\end{figure*}
\begin{figure*}
	\includegraphics[width=0.85\linewidth]{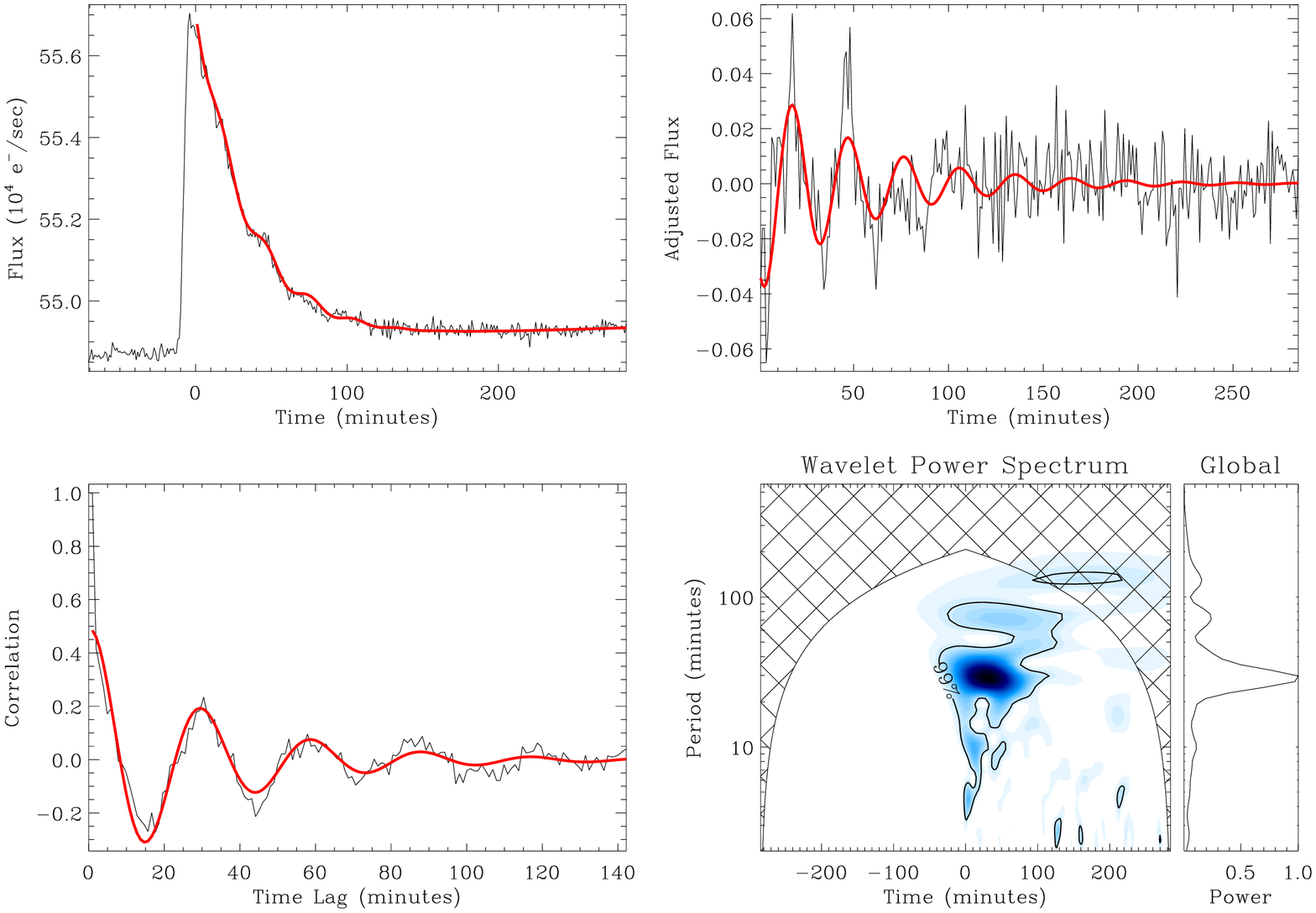}
    \caption{KIC 9946017, start time (MJD): 55217.57.}
    \label{fig:flare10}
\end{figure*}
\begin{figure*}
	\includegraphics[width=0.85\linewidth]{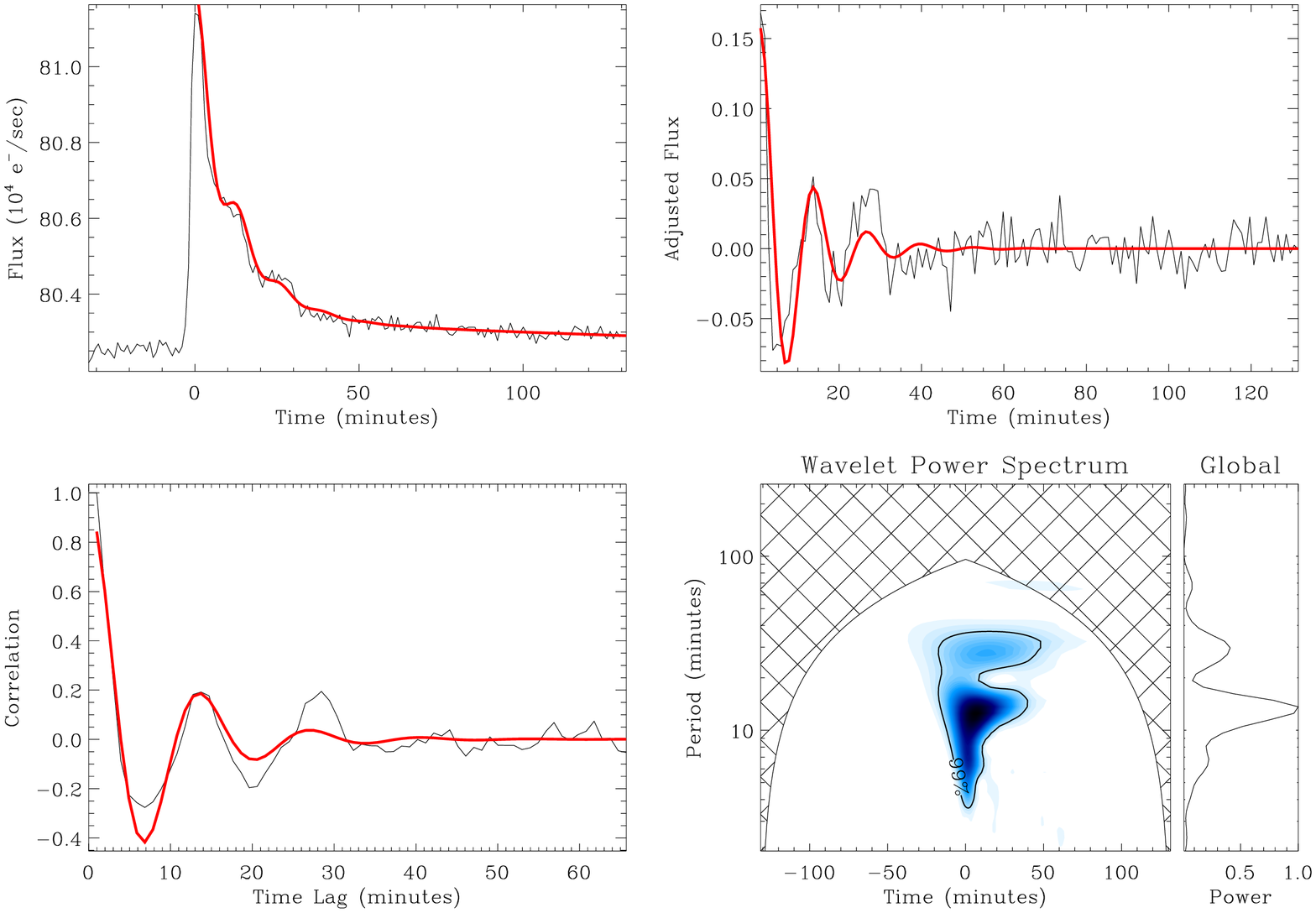}
    \caption{KIC 10459987, start time (MJD): 55158.15.}
    \label{fig:flare10}
\end{figure*}

\section{Additional tables}
\label{sec:C}

Tables containing all fit parameters for each of the flares are included in this section.

\clearpage

\onecolumn
\begin{center}
\begin{longtable}{ c c c c c c c }
\caption{Flare decay fit parameters of the flares showing evidence of QPPs. The KIC number of the star is given along with the approximate time at which the flare occurs and the fit parameters, as described by Equations \ref{eq:flarelin} and \ref{eq:flarequad}.} \\
\hline
KIC        &Time (Modified  &Amplitude,                    & Decay time,  &B (10$^{-4}$ e$^{-}$sec$^{-1}$min$^{-1}$     &C                        &D   \\
	   &Julian Date)    &$A_0$ (10$^{-4}$ e$^-$/sec)   &$t_0$ (min)   &or 10$^{-4}$ e$^{-}$sec$^{-1}$min$^{-2}$)    &(10$^{-4}$ e$^{-}$/sec)  &(min)   \\
\hline \\
\endfirsthead
\multicolumn{7}{c}%
{\tablename\ \thetable\ -- \textit{Continued from previous page}} \\
\hline
KIC        &Time       &Amplitude  &Decay time   &B                      &C      &D   \\
\hline \\
\endhead
\hline \multicolumn{7}{r}{\textit{Continued on next page}} \\
\endfoot
\hline
\endlastfoot
2852961	   &55238.22   & 0.76      &241.556      & 0.00                  &128.79 & -    \\[3pt]
2852961    &55240.27   & 7.96      &460.718      & 0.00                  &127.25 & -    \\[3pt]
3128488    &54990.32   & 1.20      & 25.132      &$-4.84 \times 10^{-4}$ & 25.68 & -    \\[3pt]
3540728    &55751.38   & 0.15      &  4.547      &$2.23 \times 10^{-6}$  & 11.39 &-283.47  \\[3pt]
3540728    &55807.25   & 1.16      & 39.078      & 0.00                  & 11.29 & -    \\[3pt]
4671547    &55090.04   & 0.52      & 18.065      & 0.00                  & 38.61 & -    \\[3pt]
4758595    &56219.15   & 0.71      & 21.292      &$-5.58 \times 10^{-4}$ & 16.54 & -    \\[3pt]
5475645    &55095.92   & 0.23      & 27.553      &$-5.23 \times 10^{-4}$ & 37.81 & -    \\[3pt]
5475645    &56330.43   & 0.24      & 87.865      &$-2.81 \times 10^{-5}$ & 40.14 & -    \\[3pt]
6184894    &56243.87   & 0.07      & 28.996      &$3.87 \times 10^{-7}$  &  7.29 &-228.22  \\[3pt]
6184894    &56291.77   & 0.17      & 10.913      &$5.61 \times 10^{-7}$  &  7.34 &-667.77  \\[3pt]
6437385    &55391.10   & 0.71      &152.034      &$-7.79 \times 10^{-6}$ & 38.84 & -   \\[3pt]
6437385    &55393.76   & 0.67      & 16.882      &$5.47 \times 10^{-6}$  & 37.69 &-346.91  \\[3pt]
7664485    &56107.70   & 0.05      & 79.561      &$-2.03 \times 10^{-5}$ &  5.73 & -   \\[3pt]
7664485    &56119.79   & 0.17      & 11.105      &$6.95 \times 10^{-7}$  &  5.72 &-252.56  \\[3pt]
7885570    &55010.88   & 0.28      & 20.804      &$-4.06 \times 10^{-6}$ & 32.32 & -82.97  \\[3pt]
7940533    &55317.42   & 0.22      & 47.852      & 0.00                  & 10.10 & -    \\[3pt]
8226464    &55012.10   & 0.49      & 50.559      &$-5.95 \times 10^{-4}$ & 34.20 & -    \\[3pt]
8414845    &56217.91   & 0.09      & 46.991      &$-6.53 \times 10^{-7}$ &  6.28 &-147.58  \\[3pt]
8414845    &56285.43   & 0.08      &118.071      &$3.20 \times 10^{-7}$  &  6.14 &-601.68  \\[3pt]
8414845    &56293.70   & 0.09      & 38.782      &$2.44 \times 10^{-4}$  &  6.12 & -    \\[3pt]
8915957    &55152.31   & 0.15      & 53.075      &$-7.52 \times 10^{-5}$ & 52.50 & -    \\[3pt]
9641031    &55614.55   & 3.70      & 19.189      &$-1.13 \times 10^{-3}$ &309.36 & -    \\[3pt]
9652680    &55085.13   & 1.48      & 21.007      &$-1.34 \times 10^{-6}$ & 45.88 & -89.16  \\[3pt]
9655129	   &56149.04   & 0.04      & 43.862      &$-7.25 \times 10^{-5}$ &  3.86 & -    \\[3pt] 
9726699	   &55382.78   & 1.02      & 11.445      &$-3.12 \times 10^{-3}$ & 27.36 & -    \\[3pt]
9726699	   &55401.16   & 0.13      &  6.317      &$-9.19 \times 10^{-6}$ & 26.55 & -37.87  \\[3pt]
9726699    &55409.48   & 0.09      & 26.260      &$-5.26 \times 10^{-6}$ & 26.47 & 118.58  \\[3pt]
9726699    &55749.56   & 0.48      & 14.658      &$-1.82 \times 10^{-3}$ & 26.57 & -    \\[3pt]
9726699    &55999.77   &14.66      & 23.487      &$-1.38 \times 10^{-2}$ & 29.69 & -    \\[3pt]
9726699    &56082.84   & 0.37      &  5.486      &$-3.66 \times 10^{-3}$ & 26.21 & -    \\[3pt]
9821078    &55487.25   & 0.10      & 13.639      &$-3.91 \times 10^{-5}$ &  3.14 & -    \\[3pt]
9946017    &55217.57   & 0.79      & 36.529      &$1.73 \times 10^{-4}$  & 54.89 & -    \\[3pt]
10206340   &55076.74   & 1.97      &128.953      &$-1.49 \times 10^{-5}$ & 60.20 &-335.74  \\[3pt]
10459987   &55158.15   & 0.72      & 12.437      &$-3.26 \times 10^{-4}$ & 80.33 & -    \\[3pt]
10459987   &56189.39   & 0.26      &  7.122      &$-8.09 \times 10^{-4}$ & 78.42 & -    \\[3pt]
10528093   &56214.53   & 0.07      & 59.488      & 0.00                  &  4.84 & -    \\[3pt]
10528093   &56262.77   & 0.25      & 54.927      &$-4.33 \times 10^{-5}$ &  4.82 & -    \\[3pt]
11551430   &55004.60   & 1.04      & 34.201      &$-3.28 \times 10^{-4}$ & 85.27 & -    \\[3pt]
11551430   &55024.13   & 1.24      & 71.061      & 0.00                  & 83.92 & -    \\[3pt]
11551430   &55031.05   & 2.71      & 63.622      & 0.00                  & 85.21 & -    \\[3pt]
11551430   &55031.96   & 1.38      & 34.188      &$-4.11 \times 10^{-4}$ & 84.91 & -    \\[3pt]
11551430   &56117.13   & 0.71      &135.828      &$-8.01 \times 10^{-4}$ & 79.85 & -    \\[3pt]
11551430   &56134.74   & 1.44      & 66.907      &$-5.11 \times 10^{-4}$ & 78.16 & -    \\[3pt]
11551430   &56166.63   & 1.39      & 55.898      &$-1.26 \times 10^{-6}$ & 79.69 &-251.91  \\[3pt]
11551430   &56208.35   & 3.94      & 50.695      &$-7.49 \times 10^{-4}$ & 79.84 & -    \\[3pt]
11551430   &56264.09   & 2.14      &109.713      & 0.00                  & 78.66 & -    \\[3pt]
11551430   &56270.75   & 6.29      & 40.643      &$-1.10 \times 10^{-3}$ & 79.68 & -    \\[3pt]
11560431   &56150.68   & 0.53      & 12.627      &$2.51 \times 10^{-4}$  &202.90 & -    \\[3pt]
11560431   &56193.12   & 0.72      & 13.623      &$-1.32 \times 10^{-3}$ &201.47 & -    \\[3pt]
11560447   &55947.44   & 3.84      &  4.803      &$-2.01 \times 10^{-2}$ & 72.76 & -    \\[3pt]
11610797   &54981.63   & 0.95      & 54.170      & 0.00                  & 33.65 & -    \\[3pt]
11665620   &55762.95   & 0.17      & 17.923      &$4.49 \times 10^{-6}$  &  2.66 &-188.60  \\[3pt]
12102573   &55086.03   & 0.02      &  3.932      &$3.63 \times 10^{-6}$  & 23.03 &-217.59  \\[3pt]
12156549   &55287.92   & 0.08      &111.846      &$3.66 \times 10^{-5}$  &  0.54 & -    \\[3pt]
12156549   &55347.20   & 0.06      & 66.821      & 0.00                  &  0.52 & -    \\[3pt]
\label{tab:flarefitparams}
\end{longtable}
\end{center}
\clearpage

\begin{table}
\centering
\caption{QPP fit parameters of the 11 flares showing evidence of stable decaying oscillations. The KIC number of the star is given along with the approximate time at which the flare occurs and the fit parameters, as described by Equations \ref{eq:qppexp} and \ref{eq:qppgau}.}
\label{tab:qppfitparams}
\begin{tabular}{c c c c c c c c}
\hline
KIC        &Time (Modified  &Amplitude,                  &Decay time,                     &B (min             &Period,           &Phase,            &Decay         \\ 
	   &Julian date)    &$A$ (10$^{-4}$ e$^-$/sec)   &$\tau_{e}$ or $\tau_{g}$ (min)  &or min$^{-2}$)     &$P$ (min)         &$\phi$            &profile       \\ 
\hline	\\
2852961	   &55238.22        &$0.146 \pm 0.006  $         &$27 \pm 2     $                 &$ 35 \pm 2    $    &$67 \pm 1      $  &$1.64 \pm 0.08  $ &Gaussian      \\[3pt] 
3540728    &55807.25        &$0.091 \pm 0.003  $         &$14.4 \pm 0.7 $                 &$ 39.0 \pm 0.4$    &$37.1 \pm 0.4  $  &$5.27 \pm 0.08  $ &Gaussian      \\[3pt] 
5475645    &55095.92        &$0.026 \pm 0.005  $         &$9 \pm 2      $                 &$ 23 \pm 2    $    &$16.2 \pm 0.9  $  &$3.8 \pm 0.6    $ &Gaussian      \\[3pt]
6184894    &56243.87        &$0.0051 \pm 0.0008$         &$59 \pm 8     $                 &$165 \pm 14   $    &$57 \pm 1      $  &$5.0 \pm 0.5    $ &Gaussian      \\[3pt] 
6437385    &55393.76        &$0.120 \pm 0.002  $         &$26.5 \pm 0.4 $                 &$ 39.6 \pm 0.7$    &$40.9 \pm 0.3  $  &$8.46 \pm 0.04  $ &Gaussian      \\[3pt] 
9655129	   &56149.04        &$0.0026 \pm 0.0003 $        &$57.5 \pm 14.6 $                &$ 2 \pm 6     $    &$78 \pm 5      $  &$3.7 \pm 0.3    $ &Exponential   \\[3pt]
9655129    &56149.04        &$0.0010 \pm 0.0002 $        &$65.1 \pm 36.5 $                &$ -3 \pm 12   $    &$32 \pm 1      $  &$0.5 \pm 0.4    $ &Exponential   \\[3pt]
9726699	   &55401.16        &$0.030 \pm 0.003  $         &$129 \pm 34   $                 &$-48 \pm 14   $    &$24.2 \pm 0.1  $  &$0.1 \pm 0.1    $ &Exponential   \\[3pt] 
9726699	   &55999.77        &$2.261 \pm 0.016  $         &$8.89 \pm 0.05$                 &$ -3.6 \pm 0.1$    &$10.93 \pm 0.01$  &$6.830 \pm 0.006$ &Gaussian      \\[3pt]
9946017	   &55217.57        &$0.045 \pm 0.002  $         &$54 \pm 6     $                 &$ -6 \pm 3   $     &$29.1 \pm 0.4  $  &$2.6 \pm 0.1    $ &Exponential   \\[3pt] 
10459987   &55158.15        &$0.158 \pm 0.009  $         &$10 \pm 1     $                 &$ -0.1 \pm 0.6$    &$12.9 \pm 0.3  $  &$6.2 \pm 0.1    $ &Exponential   \\[3pt] 
12156549   &55287.92        &$0.0103 \pm 0.0004$         &$36 \pm 2     $                 &$ 28 \pm 2    $    &$44.6 \pm 0.6  $  &$0.37 \pm 0.07  $ &Gaussian      \\[3pt]
\hline
\end{tabular}
\end{table}


\bsp	
\label{lastpage}
\end{document}